\documentclass[11pt]{article}
\usepackage[a4paper,hmarginratio=1:1,vmarginratio=2:3,
  totalwidth=15.1cm,totalheight=23.1cm]{geometry} 
\usepackage{bm,epstopdf,epsfig,amsmath,amssymb,amsfonts,colordvi,
latexsym,comment,cancel,verbatim,url}
\usepackage{slashed}
\usepackage{graphicx,graphics}
\usepackage[font=md,captionskip=8pt]{subfig}
\usepackage[usenames,dvipsnames]{color}

\usepackage{setspace}

\newcommand{\cL}{{\cal L}}

\newcommand{\cO}{{\cal O}}   

\newcommand{\cR}{{\cal R}}

\newcommand{\cW}{{\cal W}}

\newcommand{\Tr}{\mbox{Tr}}

\renewcommand{\Re}{\text{Re}\ }
\renewcommand{\Im}{\text{Im}\ }

\newcommand{\ra}{\rightarrow}
\newcommand{\be}{\begin{equation}}
\newcommand{\ee}{\end{equation}}
\newcommand{\bea}{\begin{eqnarray}}
\newcommand{\eea}{\end{eqnarray}}

\newcommand{\baa}{\begin{array}}
\newcommand{\eaa}{\end{array}}
\long\def\symbolfootnote[#1]#2{\begingroup
\def\thefootnote{\fnsymbol{footnote}}\footnote[#1]{#2}\endgroup}
\begin{document} 
\begin{flushright}
CERN-PH-TH-2016-140\\
\end{flushright}

\thispagestyle{empty}

\vspace{3.5cm}

\begin{center}
{\Large {\bf Bounds on supersymmetric effective operators

\bigskip
from heavy diphoton searches}}

\vspace{1cm}

{\bf D. M. Ghilencea$^{\,a,\,b}$} and {\bf  Hyun Min Lee$^{\,c}$ }

\bigskip

{\small  $^a$ Theory Division, CERN, 1211 Geneva 23, Switzerland}

{\small $^b$ Theoretical Physics Department, National Institute of Physics}

{\small and Nuclear\, Engineering \, (IFIN-HH)\, Bucharest\, 077125, Romania}

{\small $^c$ Department of Physics, Chung-Ang University, 06974 Seoul, Korea.}
\end{center}

\bigskip

\begin{abstract}
We identify  the  bounds on supersymmetric  effective operators beyond MSSM, 
 from  heavy diphoton resonance ($X$) negative searches at the LHC,
where $X$ is identified with the neutral CP-even (odd)  $H$  ($A$) or  both (mass degenerate).
While minimal supersymmetric models (MSSM, etc) comply with the data, 
 a leading effective operator of $d=6$ can 
 contribute significantly to diphoton production $\sigma\sim 1$ fb, 
well above its MSSM value and in conflict with   recent  data.
Both the $b\bar b$ and $gg$ production mechanisms of $H$ and $A$ can contribute comparably
 to this. We  examine the dependence of  the diphoton  cross section $\sigma$
on  the values of $m_X$,  $\Lambda$ and $\tan\beta$, 
under the  experimental constraints from the SM-like higgs couplings $hgg$ 
and $h\gamma\gamma$ (due to mixing) and from the $b\bar b$ and $t\bar t$ searches.
These give $\Lambda$ larger than $\sim  5$  TeV 
for  $m_X$ in the range $0.5-1$ TeV.
We show how to generate the $d=6$ effective operator from  microscopic (renormalizable) models.
This demands the presence of  vector-like states beyond the MSSM spectrum
 (and eventually but  not necessarily  a  gauge singlet), of mass near 
$\Lambda$ and thus outside the LHC reach.
\end{abstract}

\newpage
\section{Motivation}

Current searches for new physics at the LHC  bring increasingly strong constraints
on the parameter space of  supersymmetric models. 
Consider for example  a final diphoton state at the LHC. Then
at the parton level, the exchange of a state $X$ of spin $J$, mass $m_X$ and width $\Gamma_X$
has a cross section 
\bea\label{sigma}
\sigma(pp\ra X\ra \gamma\gamma)=\frac{2 J+1}{s \,{m_X}} \Big[ \sum_{p} C_{p\bar p} \,
 \Gamma(X\ra p\bar p)\Big]\,\frac{\Gamma(X\ra \gamma\gamma)}{\Gamma_X}
\eea
where the sum is over partons $p=\{g,b,c,s,u,d,\gamma\}$.
$C_{p\bar p}$ are  partonic integrals 
coefficients evaluated at  ${m_X}$. 
LHC searches for  a heavy diphoton resonance ($X$) 
can impact on model building beyond the Standard Model (SM), in particular 
on supersymmetric models.

Much interest was raised by the initial claim by
ATLAS and CMS Collaborations at $\sqrt s\!=\!13$ TeV  of a possible
diphoton final state of $m_X=750$ GeV 
with an excess relative to the SM \cite{750} (also \cite{atlas,cms}),
with
$\sigma(pp \ra \gamma\gamma)_{\textsf{ATLAS}}=10\pm 3 \,\textsf{fb}$ and
$\sigma(pp \ra \gamma\gamma)_{\textsf{CMS}}= 6\pm 3 \,\textsf{fb}.$
Further, the analysis of additional data invalidated this claim  \cite{ncms}.
This is actually  welcome for  minimal supersymmetric models (MSSM, etc) where
 it is not possible to have a heavy resonance $X$  with such significant $\sigma$  \cite{DK},
except if\footnote{
Many non-supersymmetric  explanations were reported for this 750 GeV ``resonance'', see \cite{strumia}
for a long list of references. 
 $X$ was a  scalar singlet  with couplings to new TeV states that mediate (at loop level) its 
 production by  $gg$ fusion  and its decay to $\gamma\gamma$  \cite{GG,AA}.
For a  non-supersymmetric effective study see \cite{Zupan,Zupan2}.}.  
a): one  is  fine-tuning the parameters  \cite{DA1}
with $X$   the CP even/odd heavy higgs $X=H$, $A$,  or b): considers
 the rather  special case of  low, TeV-scale  supersymmetry breaking
with $X$ a sgoldstino \cite{Demidov}, see also  \cite{VL,UE,osusy}.

However, we show that  effective operators beyond the  MSSM (minimal) higgs sector
can contribute dramatically  to the  diphoton production (giving $\sigma\sim $ few fb)
not seen in the  data \cite{ncms}.
 The resonance $X$ is  the CP-odd/even neutral MSSM higgs $A$ or $H$ {\it  or both} (mass degenerate). 
This result  is  due to enhanced couplings of the higgs sector to SM gauge bosons,
induced by the following  {\it unique, leading}  operator of  dimension $d\!=\!6$
\bea\label{op6}
(1/\Lambda^2)\int d^2\theta \,(H_1.H_2)\, \Tr\, W^\alpha W_\alpha+\rm{h.c.},
\eea 
where  $W_\alpha$ is  the supersymmetric field strength of the SM sub-groups
U(1)$_Y$, SU(2)$_L$, SU(3).

Depending on $\Lambda$, operator (\ref{op6})
 can bring  a large correction to  the diphoton production 
in conflict with the latest data, with  impact on $H,A$ searches.
Motivated by this, we study the constraints on this operator and
 examine the dependence of the diphoton  cross section $\sigma$
on the values of $m_X$,  $\Lambda$ and $\tan\beta$,  while including
both  the  $b\bar b$  and $gg$ production   mechanisms of $X=A,H$.
The  experimental constraints on the SM-like higgs ($h$) couplings $hgg$ and $h\gamma\gamma$
and on the $b\bar b$ ($t\bar t$) cross section (that receive corrections from (\ref{op6})),
are also applied, with impact on the allowed  $m_X$,  $\Lambda$ and $\tan\beta$. 
For a given $\sigma\sim 0.1 - 1$ fb, we illustrate  these constraints for $m_{H,A}$ in 
the range $0.5-1$ TeV (in particular for the  absent ``resonance'' at  $750$ GeV).   
We then show how operator (\ref{op6}) is generated in a renormalizable model
beyond MSSM;  an extra   $d\!=\!5$ operator may also be generated (in some cases) and
 does not directly affect the diphoton production but may improve
naturalness~\cite{ft}.

In  section~\ref{2} we study the new 
couplings induced  in the MSSM higgs sector by the $d\!=\!6$  effective operator and its effect
on the diphoton production.  Section~\ref{higgs} shows 
how these  operators  are generated in a renormalizable model. Our conclusions 
are found in Section~\ref{conc}.

\section{Effective operators and diphoton resonance}\label{2}

We consider the MSSM model extended by (supersymmetric) effective operators in the higgs sector
and study the  diphoton production cross section $\sigma$ due to a  possible  resonance $X$  identified with 
 $H$ and/or $A$. We compute the corrections to the couplings of
 $H$, $A$ and $h$, and the branching ratios of $H$, $A$. We  then illustrate  
the correlations between the values of  $\Lambda$,  $m_X$, $\tan\beta$ and $\sigma$, 
consistent with  the constraints from Higgs signals/decays.

\subsection{New couplings from effective operators beyond MSSM}\label{eff}

 From all effective operators of dimensions $d=5$ and $d=6$ \cite{ft,P,d=6} beyond the MSSM
higgs sector,  we find  only one leading operator that can  contribute to  a diphoton resonance 
\bea
\cL_j &=&
 \frac{c_j}{2 \Lambda^2\, g^2_j \kappa_j}
\int d^{2}\theta 
\,\,(H_2.H_1) \,\,{\rm Tr}\,(W^\alpha W_\alpha)_j +\mbox{h.c.} 
\label{dim6}
\eea

\medskip\noindent
which has dimension $d=6$. Here $j=1,2,3$ labels the U(1)$_Y$, SU(2)$_L$, SU(3)  gauge groups of gauge 
couplings $g_j$, so we actually have three operators, with 
coefficients\footnote{The coefficients $c_j=\cO(1)$, $j=1,2,3$
 enable us later to turn on/off any of operators $\cL_{1,2,3}$.}
 $c_j=\cO(1)$ and
$\Lambda$ a free parameter.
 $\kappa_j$ is  a constant that  cancels the trace factor. 
$(W^\alpha)_j$ is the field strength of a vector superfield $V_j$. 
 The relevant part  is\footnote{Notation used: $h_1.h_2=h_1^0 h_2^0-h_1^- h_2^+$;
also $\Re h_1^0=H\,\cos\alpha  -h\,\sin\alpha$, $\Re h_2^0=H\,\sin\alpha +h\,\cos\alpha$.}
\bea
\cL_j\supset 
  \frac{c_j}{\Lambda^2}\,\, \Big[\,(h_1 . h_2)\,
(F_j^{a\,\mu\nu}F^a_{j\,\mu\nu}+ i F^a_{j\,\mu\nu} \tilde F^{a\,\mu\nu}_{j}) +\mbox{h.c.}\Big]
\eea

 \medskip\noindent
with  $\tilde F^{\mu\nu}=(1/2)\, \epsilon^{\mu\nu\rho\sigma}\, F_{\rho\sigma}$.
With real $c_j$ one has, in a standard notation
\medskip
\bea\label{ss2}
\!\!\cL_1\!+\!\cL_2\!\!\!&\supset&\!\!\!
\frac{v}{\Lambda^2}
\Big[(c_{\gamma\gamma} h  + b_{\gamma\gamma}  H )\, F_{\mu\nu} F^{\mu\nu}
+ 
(c_{\gamma z} h  + b_{\gamma z} H) \, F_{\mu\nu} Z^{\mu\nu}
\nonumber\\[5pt]
&&\quad+\,
(c_{zz} h + b_{zz} H)\, Z_{\mu\nu} Z^{\mu\nu}
+
(c_{ww}h+ b_{ww}H)\, W^+_{\mu\nu} W^{- \mu\nu}
\nonumber\\[5pt]
&&\quad+\,
a_{\gamma\gamma} \, A F_{\mu\nu}\tilde F^{\mu\nu}
+
a_{zz} A\,  Z_{\mu\nu} \tilde Z^{\mu\nu}
+
a_{\gamma z} A\, F_{\mu\nu} \tilde Z^{\mu\nu} 
+
a_{ww}\, A \, W^+_{\mu\nu} {\tilde W}^{- \mu\nu}
\Big]\quad
\eea

 \medskip\noindent
where $F_{\mu\nu}$ is the photon field strength,  $H$  ($A$) are the CP-even (odd) neutral higgses and
\medskip
\begin{gather}
a_{\gamma\gamma}  = - (c_1\, c_w^2+ c_2\, s_w^2),\quad\,
a_{zz} =  - (c_1 s_w^2+ c_2 c_w^2),\,\,\,\,
a_{\gamma z}=  -(c_2-c_1) s_{2w}, \,\,
a_{ww}= -2 c_2 \qquad\,\,\,\,\quad
\nonumber\\
b_{\gamma\gamma}=- a_{\gamma\gamma}\, s_{\alpha\beta}, 
\qquad\qquad
b_{zz}= - a_{zz}\, s_{\alpha\beta},\qquad\quad\,\,\,
b_{\gamma z}= - a_{\gamma z}\, s_{\alpha\beta},\qquad\,\,\,
b_{ww}= - a_{ww}\, s_{\alpha\beta}\qquad
\nonumber\\
c_{\gamma\gamma}=  - a_{\gamma\gamma}\,c_{\alpha\beta}, 
\qquad\quad\,
c_{zz}=-a_{zz}\, c_{\alpha\beta},\qquad\quad\,\,\,
c_{\gamma z}= -a_{\gamma z}\, c_{\alpha\beta},\qquad\,\,\,
c_{ww}= -a_{ww}\, c_{\alpha\beta}
\label{R12}
\end{gather}

\medskip\noindent
with the notations: 
$c_{\alpha\beta}=\cos(\alpha+\beta)$, $s_{\alpha\beta}=\sin(\alpha+\beta)$, $c_{2\beta}=\cos(2\beta)$,
$c_w=\cos\theta_w$, $s_w=\sin\theta_w$, $s_{2w}=\sin 2\theta_w$.
Let us also consider the effect of the  gluon operator\footnote{If $c_3$ has an imaginary part,
one also has
 $\cL_3\supset  v/\Lambda^2
 [\tilde a_{gg} A^0 \,\Tr G^2+  \tilde b_{gg} H \,\Tr G\tilde G +
\tilde c_{gg} h \,\Tr G^2]$
with $\tilde a_{gg} =-\Im[c_3]$, $\tilde b_{gg}=-\Im[c_3]\sin(\alpha+\beta)$, 
$\tilde c_{gg}=-\Im[c_3] \cos(\alpha+\beta)$.}
\medskip
\bea\label{R3}
\cL_3\supset
\frac{v}{\Lambda^2}
\Big[c_{gg}\, h\,{\rm Tr}\,G_{\mu\nu} G^{\mu\nu}
+
b_{gg} H\,{\rm Tr}\, G_{\mu\nu} G^{\mu\nu}
+
a_{gg} A\,{\rm Tr}\,G_{\mu\nu} \tilde G^{\mu\nu}\Big]
\eea
where\quad
\bea\label{R4}
a_{gg}=- c_3,\qquad
b_{gg}= c_3 \,s_{\alpha\beta},\qquad\quad
c_{gg}= c_3 \,c_{\alpha\beta}.
\eea
The Lagrangian of the MSSM corrected with $\cL_{1,2,3}$  induces the
following couplings
\bea
{\cal L}&=&
\frac{1}{v}\,\Big[
(\hat c_{\gamma\gamma} h  + \hat b_{\gamma\gamma}  H  )\, F_{\mu\nu} F^{\mu\nu}
+ 
(\hat c_{\gamma z} h  + \hat b_{\gamma z} H) \, F_{\mu\nu} Z^{\mu\nu}
+
(\hat c_{zz} h+ \hat b_{zz} H) Z_{\mu\nu} Z^{\mu\nu}
\nonumber\\[5pt]
&&\,\, +\,\,
(\hat c_{ww} h+ \hat b_{ww} H)\, W^+_{\mu\nu} W^{- \mu\nu}
+\,\,
\hat a_{\gamma\gamma} \, A\, F_{\mu\nu}\tilde F^{\mu\nu}
+
\hat a_{zz} A\, Z_{\mu\nu} \tilde Z^{\mu\nu}
+\,\,
\hat a_{\gamma z}\, A\, F_{\mu\nu} \tilde Z^{\mu\nu} 
\nonumber\\[5pt]
&& \,\, +\,
\hat a_{ww}\,A \,\,W^{+\mu\nu}\tilde W_{\mu\nu}^-
+ 
\big( \hat c_{gg}\,h + \hat b_{gg} H\big) \,\Tr\,G_{\mu\nu} G^{\mu\nu}
+ 
\hat a_{gg}\,A \,\, \Tr\,G_{\mu\nu} \tilde G^{\mu\nu}
\Big].\label{dd}
\qquad
\eea
The coefficients $\hat a$, $\hat b$ and $\hat c$ above  are related to
their counterparts without a hat:
\bea
\hat \tau_{\gamma\gamma}&  = &  \frac{\alpha_{\rm em}}{8\pi}\,\, \tau_{\gamma\gamma}^{loop}
+ \frac{v^2}{\Lambda^2} \, \tau_{\gamma\gamma}, 
\qquad\quad
\hat \tau_{zz}  =\frac{\alpha_{\rm em}}{8\pi}\,  \tau_{zz}^{loop}+ 
\frac{v^2}{\Lambda^2}\,\tau_{zz},
\nonumber\\
\hat \tau_{gg} &  = & \frac{\alpha_3}{12 \pi}\,\, \tau_{gg}^{loop}
+ \frac{v^2}{\Lambda^2} \, \tau_{gg}, 
\qquad\quad
\hat \tau_{ww} =  \frac{\alpha_{\rm em}}{8\pi}\, \tau_{ww}^{loop}+
\frac{v^2}{\Lambda^2}\,\tau_{ww},\,\,\,
\nonumber\\
 \hat \tau_{\gamma z} & = & 
\frac{\alpha_{\rm em}}{8\pi s_w} \, \tau_{\gamma z}^{loop}+  
\frac{v^2}{\Lambda^2} 
\, \tau_{\gamma z}, 
\qquad\quad
{\rm where}\,\,\,\tau=a, b, c.
\label{hat}
\eea

\medskip\noindent
The coefficients multiplying $v^2/\Lambda^2$   are those of  eqs.(\ref{R12}), (\ref{R4}). 
Further,  the coefficients  $a_{..}^{loop}$, $b_{..}^{loop}$, $c_{..}^{loop}$
are  loop-induced, due to the MSSM  (in the absence of the effective operators).
They bring a very small branching ratio to photons \cite{DK} relative
to $v^2/\Lambda^2$ terms and we  present them  in Appendix~A in the decoupling limit
($\alpha\ra \beta-\pi/2$)  in which we work in this paper.
We  show that their corrected version $\hat a$, $\hat b$, $\hat c$ of eqs.(\ref{dd}), (\ref{hat})
can bring a heavy diphoton resonance of large $\sigma\sim 1$ fb, in our 
model defined by MSSM extended
 by  eq.(\ref{dim6}), in  possible conflict with the latest data.

\subsection{Decay branching ratios  of $A$ and $H$}
\label{br}

\begin{figure}[t!]
\begin{center}
\includegraphics[width=6.cm,height=5.5cm]{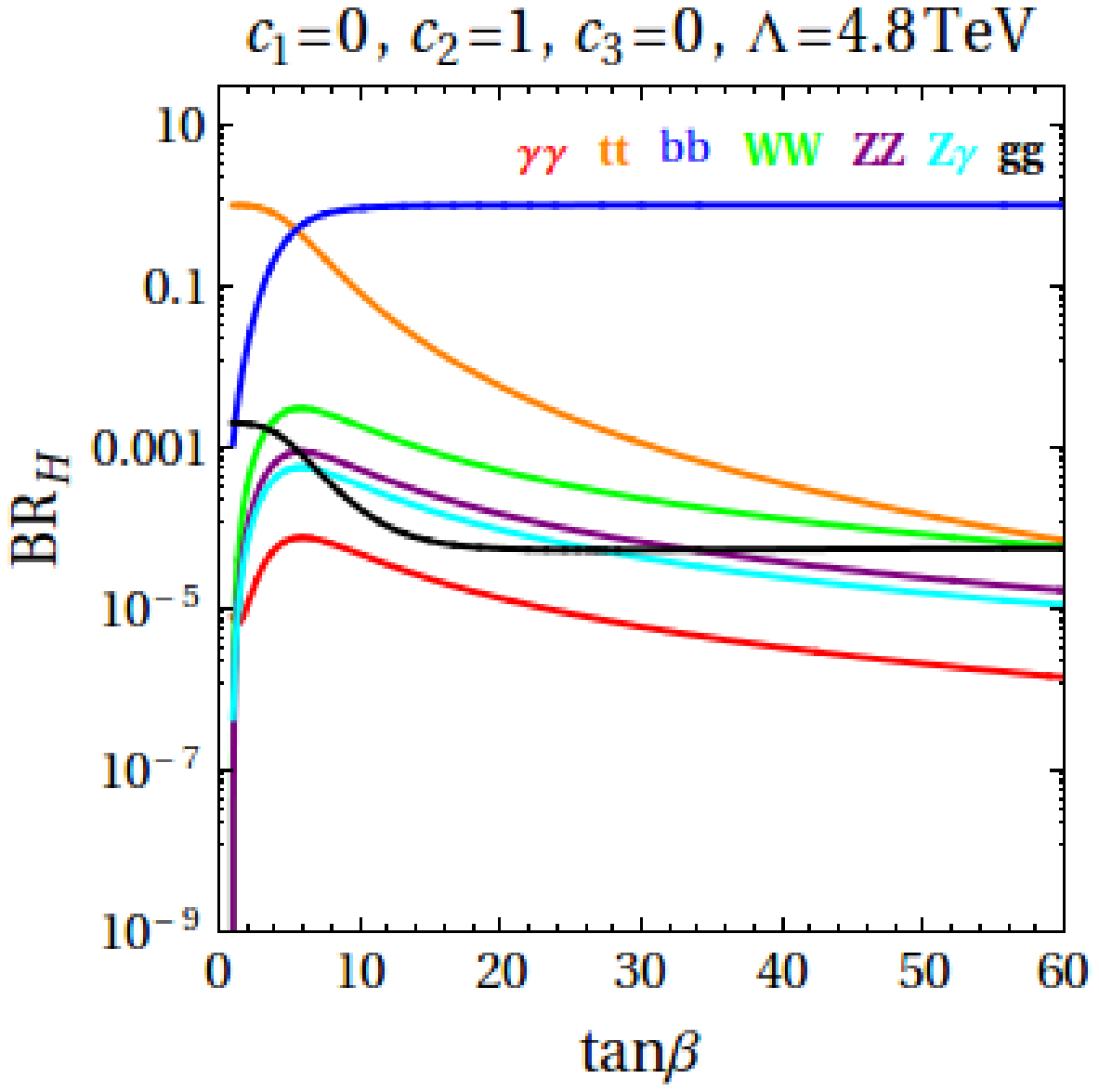}
\includegraphics[width=6.cm,height=5.5cm]{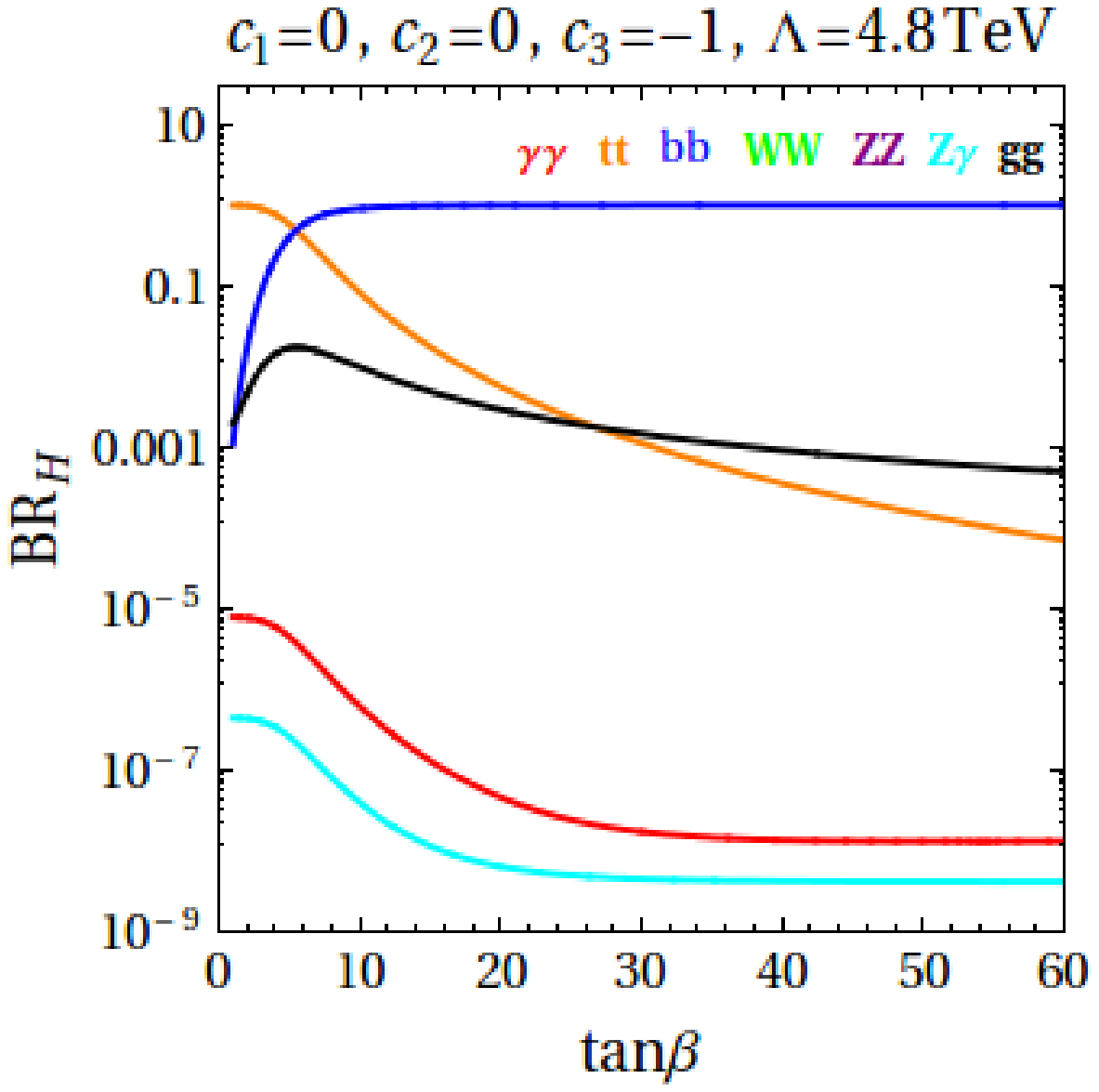}
\\[5pt]
\includegraphics[width=6.cm,height=5.5cm]{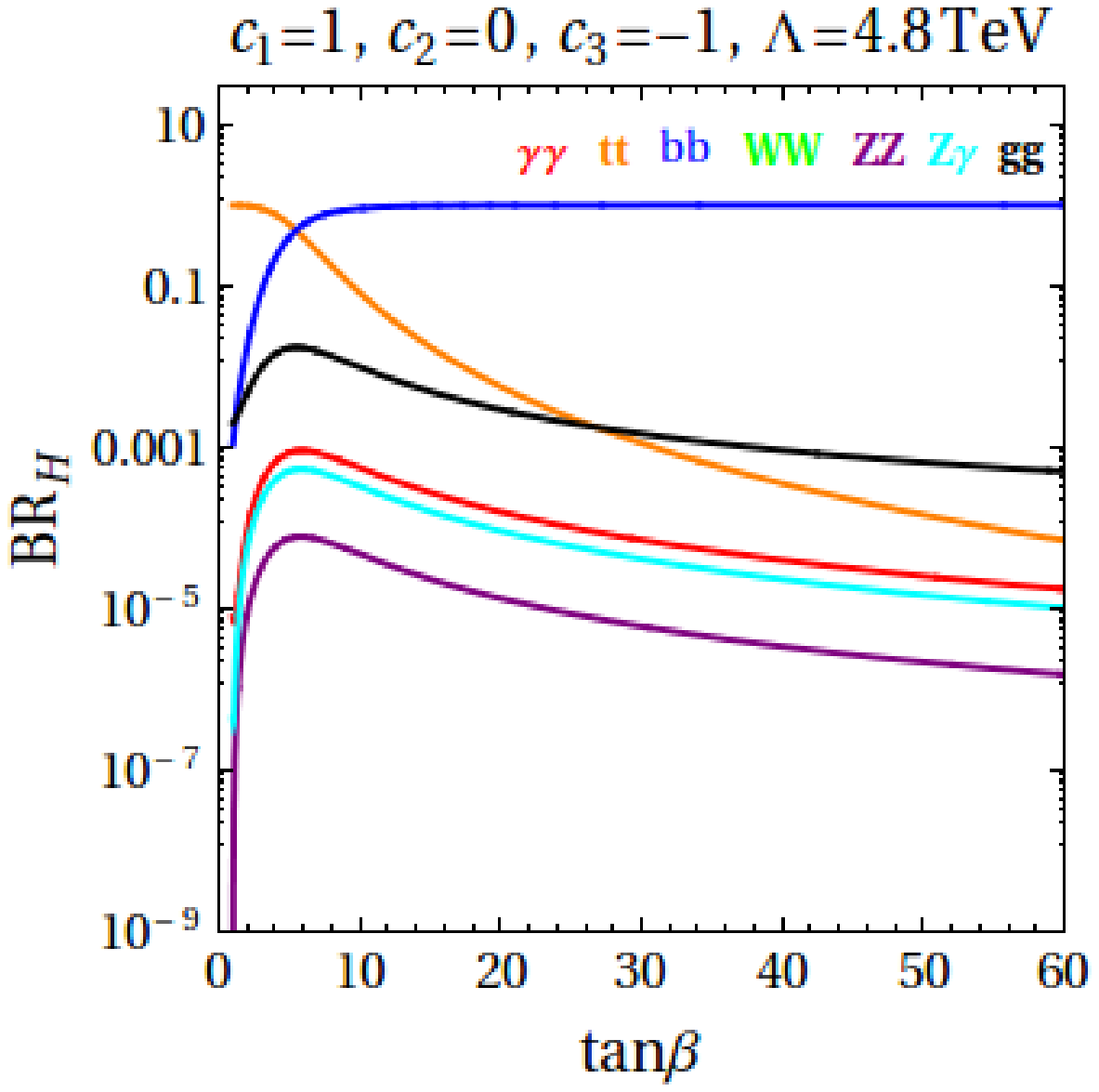}
\includegraphics[width=6.cm,height=5.5cm]{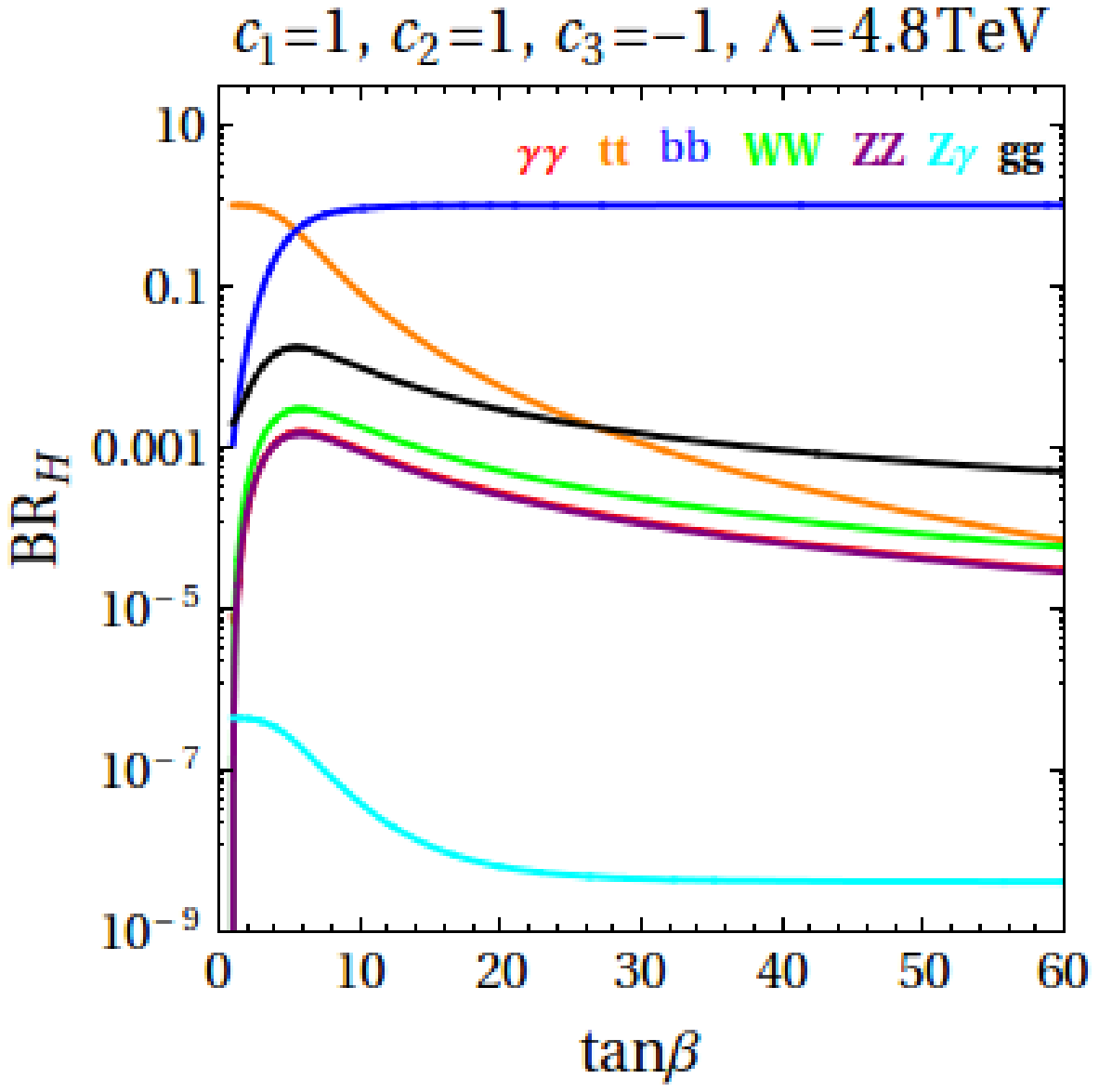}
\end{center}
\renewcommand{\baselinestretch}{0.9}
\vspace{-0.2cm}
\caption{\small
Branching ratios of $H\ra\gamma\gamma$, $t\bar t$, $b\bar b$, $WW$, $ZZ$, $Z\gamma$ 
and $gg$, for different $c_{1,2,3}$ and $\tan\beta$. The case  $c_1=1$, 
$c_2=c_3=0$ (not shown) is similar to $c_1=0$, $c_2=1$, $c_3=0$ (without $WW$).
The branching ratios for $A$ into the same final states are similar to those above,
for the same parameters. Compared to individual $\cL_{1,2,3}$, a
 combination $\cL_{1}+\cL_3$ or $\cL_1+\cL_2+\cL_3$  brings the largest branching ratio
of H (A) to $\gamma\gamma$, for appropriate relative signs of $c_j$, $j=1,2,3$.}
\label{fig1}
\end{figure}

\begin{figure}[th]
  \begin{center}
\includegraphics[width=5.7cm,height=5.5cm]{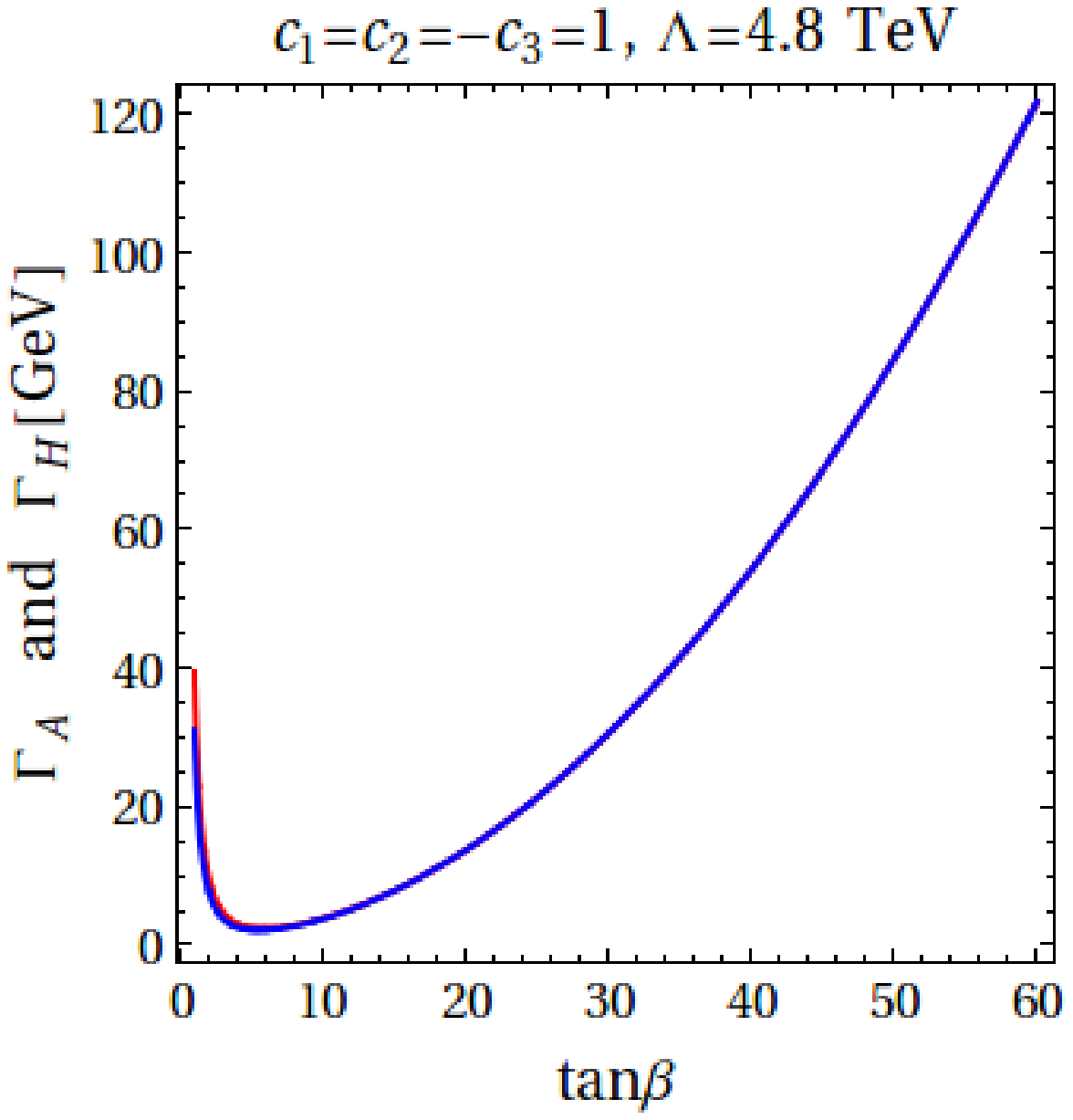}\,\,\,\,\,
\includegraphics[width=5.7cm,height=5.5cm]{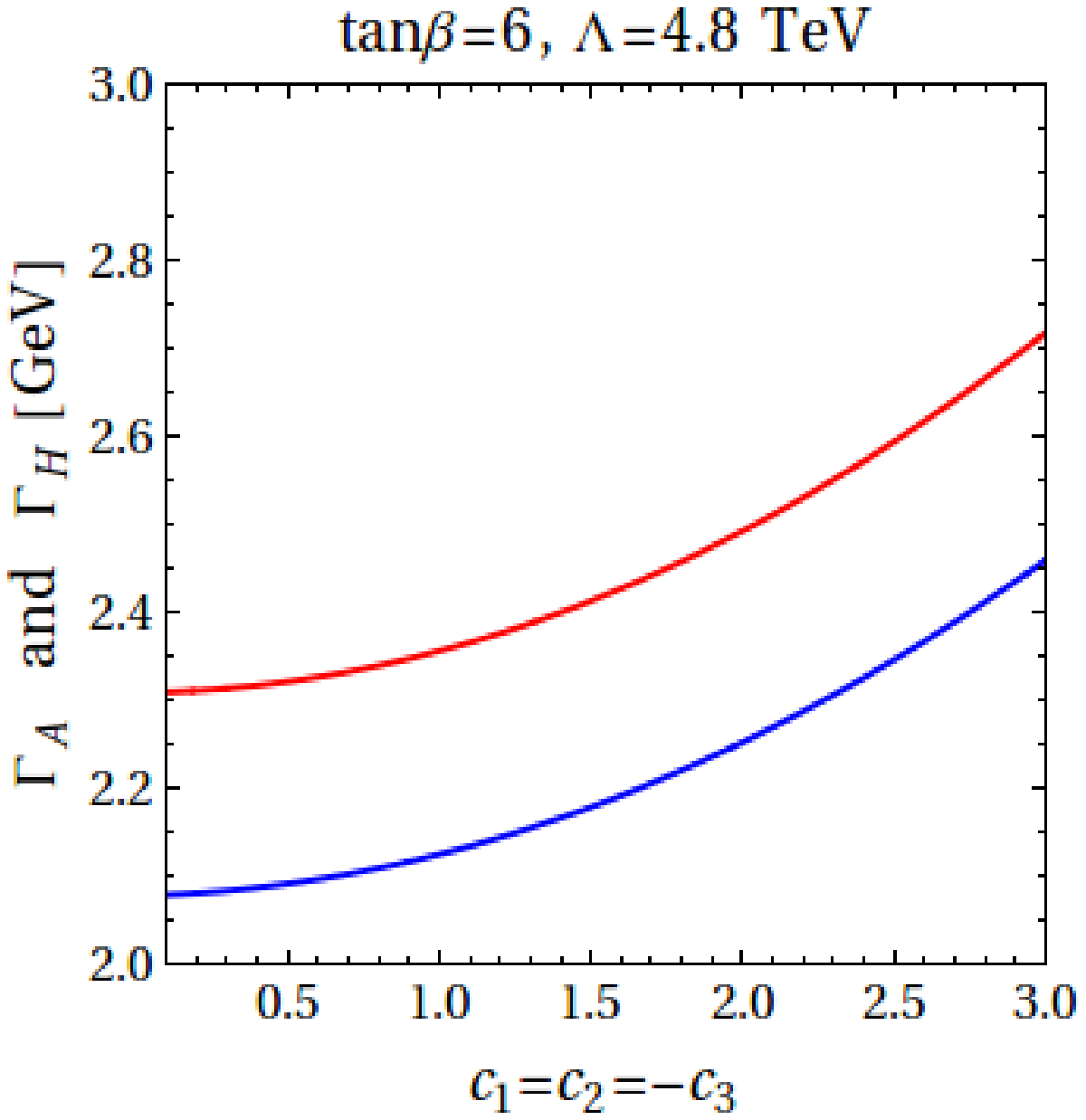}
   \end{center}
  \caption{
\small Total decay widths of heavy Higgs fields  $\Gamma_H$ (blue) and $\Gamma_A$ (red), for
 $m_{A,H}=750$ GeV. These plots remain similar for other   $c_1, c_2, c_3$ used later.
A  narrow width region $\Gamma_X\leq 5$ GeV corresponds to low $2\!\leq\! \tan\beta\!\leq\! 12$
while for larger $\tan\beta$ one has a large width regime.}
  \label{fig2}
 \end{figure}

To discuss the diphoton production we first analyze 
the  impact of the corrections in  
eq.(\ref{dd}), (\ref{hat}) on the decay rates of the heavy neutral 
CP-even (odd) Higgs $H$ ($A$), respectively.
The decay rate of $H$  is
$\Gamma_H=\sum_i \Gamma(H\ra i)$, where
\bea
\Gamma(H\ra t {\bar t}) &=& \frac{N_c m^2_t}{8\pi v^2}\,\cot^2\beta\, m_H (1-4x_t )^{3/2}, 
\nonumber\\
\Gamma(H\ra b{\bar b})&=&\frac{N_c m^2_b}{8\pi v^2}\,\tan^2\beta\, m_H (1-4x_b )^{3/2},
\qquad
x_i\equiv m^2_i/m^2_H
\eea
valid in the decoupling limit and
\bea
\Gamma(H\ra \gamma\gamma)&=& \frac{1}{4\pi} \Big(\frac{\hat b_{\gamma\gamma}}{v}\Big)^2 m_H^3
\nonumber\\[6pt]
\Gamma(H\ra\gamma Z)&=& \frac{1}{8\pi} \Big(\frac{\hat b_{\gamma z}}{v}\Big)^2 m_H^3\,(1-x_Z)^3
\nonumber\\[6pt]
\Gamma(H\ra gg)&=& \frac{2}{\pi} \,\,\Big(\frac{\hat b_{gg}}{v}\Big)^2  m^3_H 
\nonumber\\[6pt]
\Gamma(H\ra WW) &=& \frac{1}{8\pi} \Big(\frac{\hat b_{ww}}{v}\Big)^2 m^3_H (1-4x_W+6x^2_W)(1-4x_W)^{1/2},
\nonumber\\[6pt]
\Gamma(H\ra ZZ)&=&  \frac{1}{4\pi} \Big(\frac{\hat b_{zz}}{v}\Big)^2 m_H^3 (1-4x_Z+6x^2_Z)(1-4x_Z)^{1/2}.
\eea

\medskip\noindent
The decay rate of the heavy neutral CP-odd Higgs $A$
is  $\Gamma_A=\sum_i \Gamma(A\ra i)$, with
\bea
\Gamma(A\ra t {\bar t}) &=& \frac{N_c m^2_t}{8\pi v^2}\,\cot^2\beta\, m_A (1-4\bar x_t )^{1/2}, 
\nonumber\\[4pt]
\Gamma(A\ra b{\bar b})&=&\frac{N_c m^2_b}{8\pi v^2}\,\tan^2\beta\, m_A (1-4\bar x_b )^{1/2},
\qquad\qquad\qquad
\nonumber\\[4pt]
\Gamma(A\ra \gamma\gamma)&=& \frac{1}{4\pi} \Big(\frac{\hat a_{\gamma\gamma}}{v}\Big)^2  m^3_A,
\nonumber\\[4pt]
\Gamma(A\ra\gamma Z)&=& \frac{1}{8\pi} \Big( \frac{\hat a_{\gamma z}}{v} \Big)^2\,m^3_A (1-\bar x_Z)^3,
\qquad\qquad\qquad\qquad
\nonumber\\[4pt]
\Gamma(A\ra gg)&=&\frac{2}{\pi}\,\, \Big(\frac{\hat a_{gg}}{v}\Big)^2\, m^3_A 
\nonumber\\[4pt]
\Gamma(A\ra WW) &=& \frac{1}{8\pi} \Big(\frac{\hat a_{ww}}{v}\Big)^2 m^3_A \,( 1-4\bar x_W)^{3/2},
\nonumber\\[4pt]
\Gamma(A\ra ZZ)&=& \frac{1}{4\pi} \Big(\frac{\hat a_{zz}}{v}\Big)^2 m_A^3\, (1-4\bar x_Z)^{3/2},
\qquad {\bar x}_i\equiv m^2_i/m^2_A.
\eea

\medskip\noindent
In figure~\ref{fig1} the branching ratios of $H$ decays are presented  as functions of $\tan\beta$,
for different $c_{1,2,3}$. The dominant decays modes are 
into $t\bar t$ at low $\tan\beta<6$ and $b \bar b$ at large $\tan\beta$ 
while near  $\tan\beta\sim 6$ or so, they  are comparable.
The remaining decay modes have smaller, often comparable rates.
For  $A$, one has nearly identical plots. Compared to individual $\cL_{1,2,3}$,
a combination  $\cL_1+\cL_3$ or $\cL_1+\cL_2+\cL_3$  brings the
largest branching ratio of $H$ ($A$) to $\gamma\gamma$, for suitable
 relative signs of $c_{1,2,3}$ (shown). 
As an illustration, we used $m_X\!=\!750$ GeV ($X=H,A$) but these plots are  similar for 
$500\! \leq\! m_X\! \leq\! 1000$ GeV.
The total $\Gamma_{H,A}$ is shown in figure~\ref{fig2}.
  $\tan\beta$  controls the width of the resonance $X$ ($X=A$ or $H$).
At low $2\leq\! \tan\beta\!\leq\! 12$,  $\Gamma_X\!\leq\! 5$GeV and
one has the limit of narrow width  ($\Gamma_X/m_X\! \ll 1$).
Figure~\ref{fig2} remains similar for other  $c_{1,2,3}\sim \cO(1)$, of different
signs, or if $c_1$ or $c_2$ vanish.

\subsection{Diphoton searches at large $m_{H,A}$}
\label{di}

Assuming a resonance $X=H, A$, we include  the dominant $gg$ and $b\bar b$
production channels  and consider the contributions of   $A$ and $H$ to a
diphoton final state. From eq.(\ref{sigma})
\bea\label{deg}
\sigma(pp\ra \underbrace{H, A}_{X}\ra \gamma\gamma)&=& \frac{1}{sm_H}(K_{gg} C_{gg} \Gamma(H\ra gg)
+K_{b{\bar b}}C_{b{\bar b}} \Gamma(H\ra b{\bar b}))\,\textsf{Br}(H\ra\gamma\gamma) \nonumber \\
&+&\frac{1}{sm_A}(K_{gg}C_{gg} \Gamma(A\ra gg)
+K_{b{\bar b}}C_{b{\bar b}} \Gamma(A\ra b{\bar b}))\, \textsf{Br}(A\ra\gamma\gamma)\,\,\,\,\quad
\eea

\begin{figure}[h!]
  \begin{center}
    \includegraphics[width=6cm,height=5.5cm]{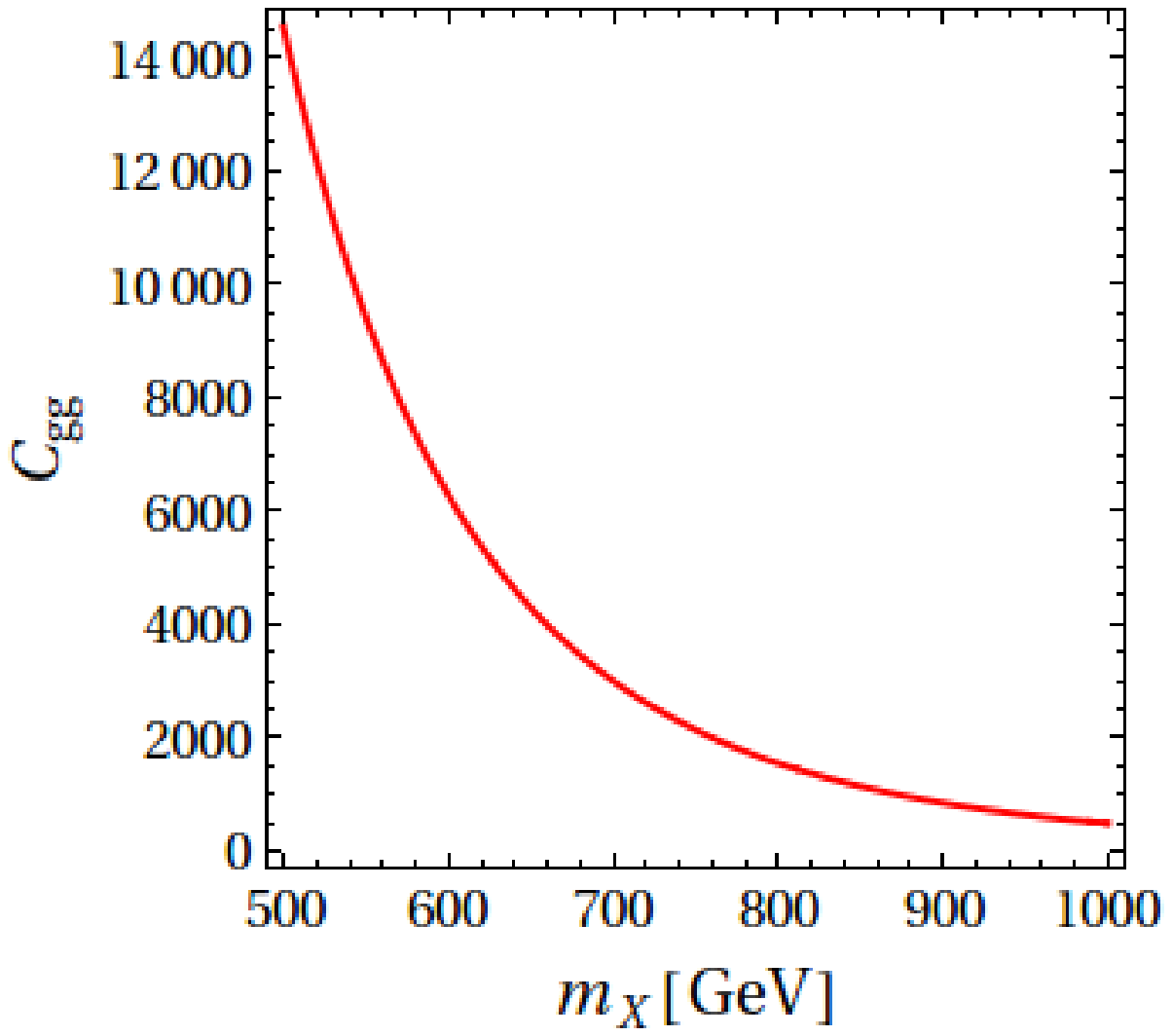} 
 \hspace{0.3cm}
 \includegraphics[width=5.8cm,height=5.45cm]{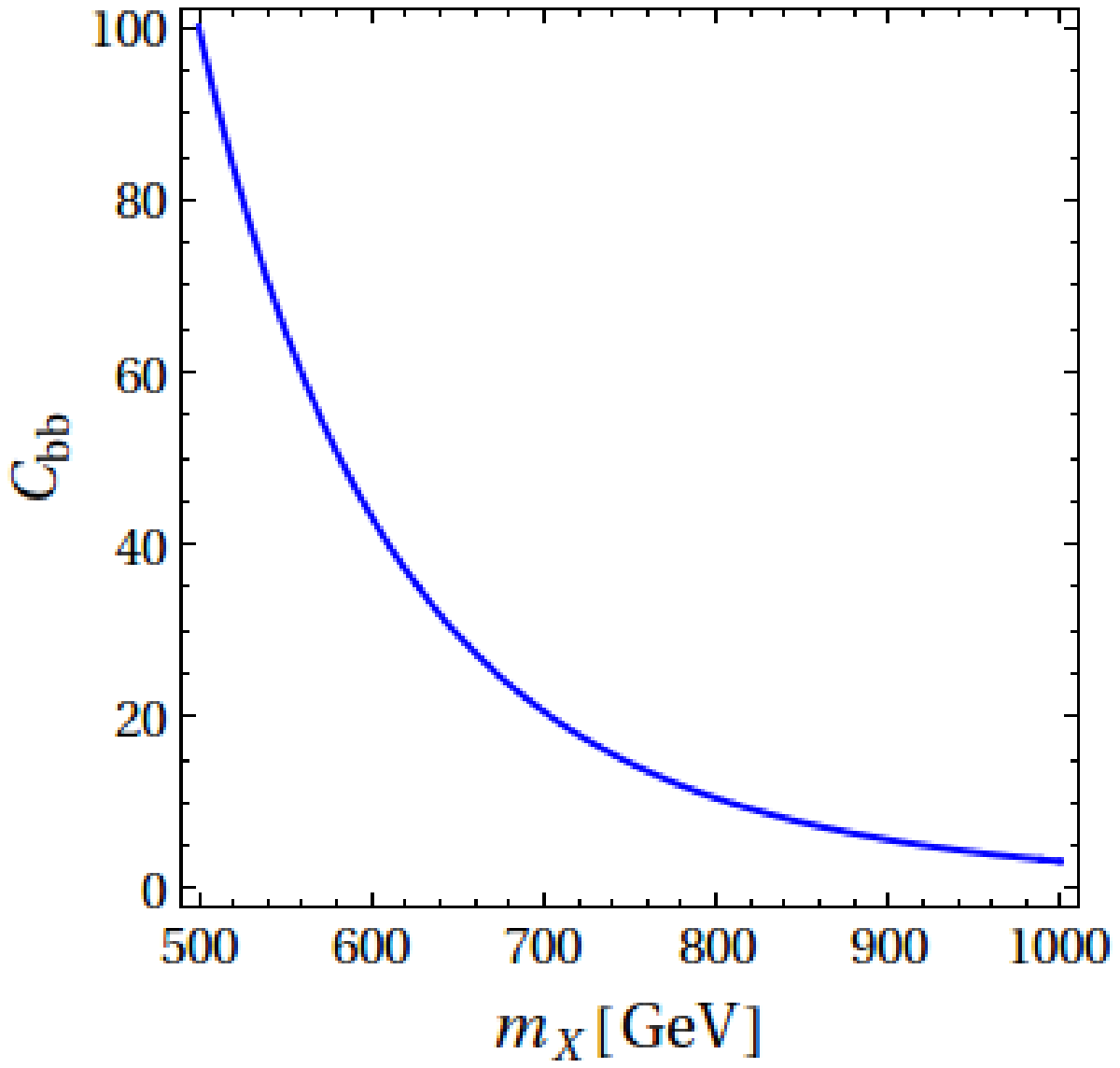} 
\end{center}
\caption{\small The dependence of  partonic integrals coefficients $C_{gg}$ and 
$C_{b\bar b}$  at $\sqrt s=13$ TeV on  $m_X$  \cite{cteq5}.
 In the model considered here  $X=H, A$.
For example, for $m_X=750$ GeV one has
  $C_{gg}=2131$, $C_{b{\bar b}}=14.6$; also
 $C_{\gamma\gamma}\approx 54$, $C_{u\bar u}\approx 1054$, $C_{d\bar d}\approx 627$,
 $C_{c\bar c}\approx 36$,  for $\sqrt s=13$ TeV.}
\label{figc}
\end{figure}

\medskip\noindent
where $K_{gg}, K_{b{\bar b}}$ are K-factors, given by $K_{gg}=1.5, K_{b{\bar b}}=1.2$, 
and $C_{gg}, C_{b{\bar b}}$ are parton luminosities.
Their values depend on the mass of the resonance, as shown in 
figure~\ref{figc}, that we generated with the CTEQ5 package \cite{cteq5}.

\begin{figure}[h!]
  \begin{center}
    \includegraphics[width=6cm,height=5.4cm]{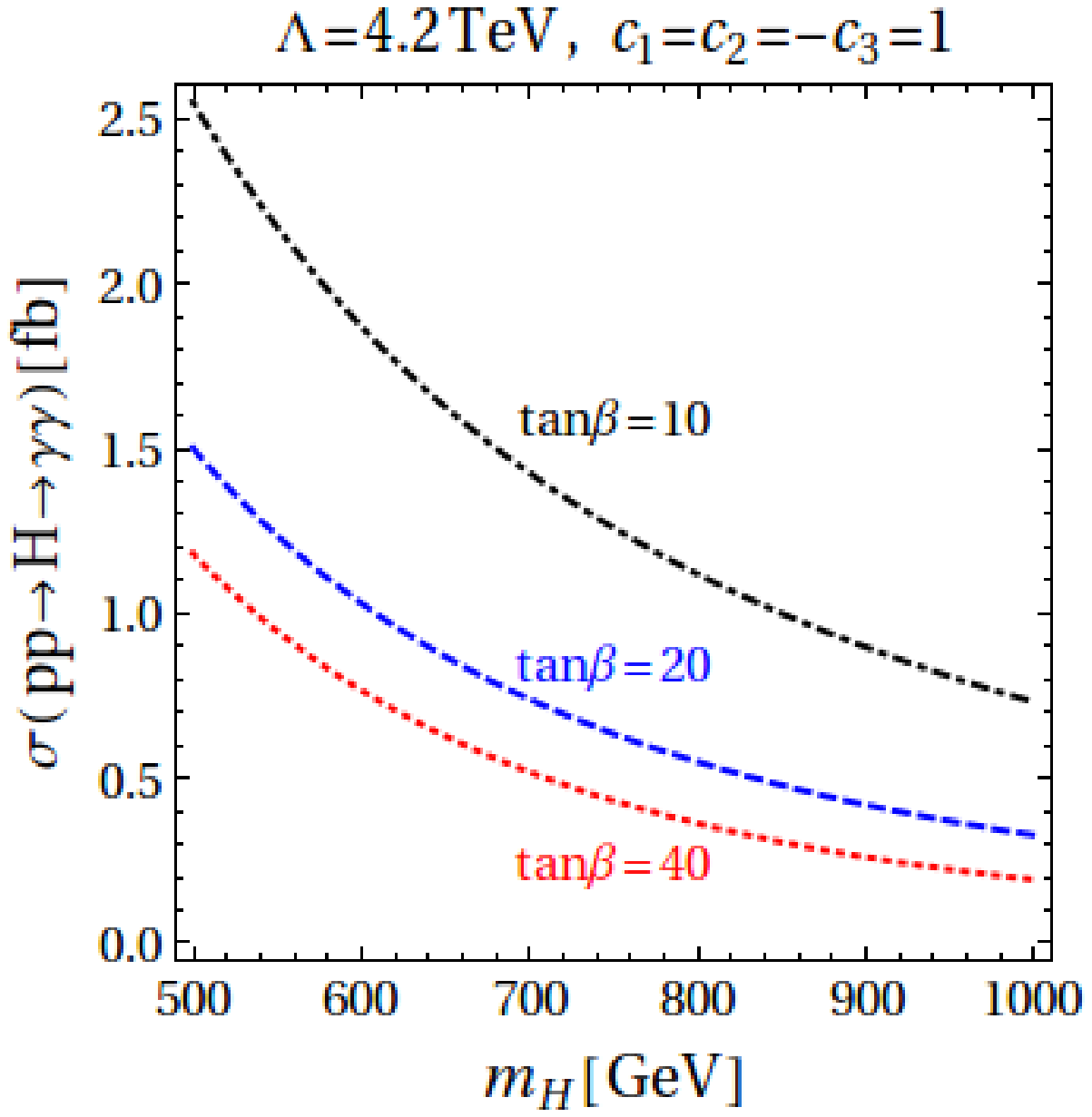} 
 \hspace{0.3cm}
\includegraphics[width=6cm,height=5.4cm]{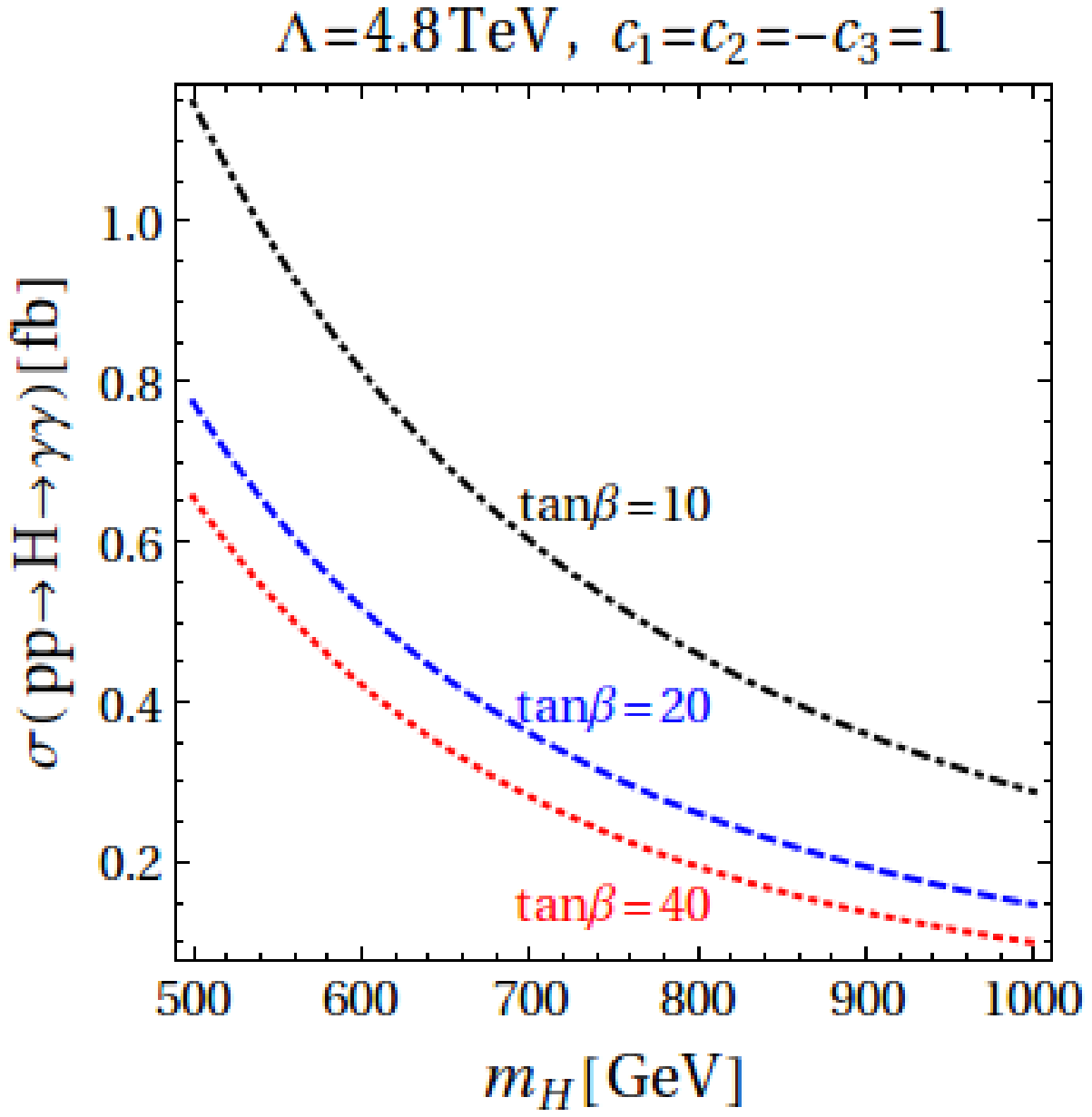} 
 \\[6pt]
    \includegraphics[width=6.1cm,height=5.4cm]{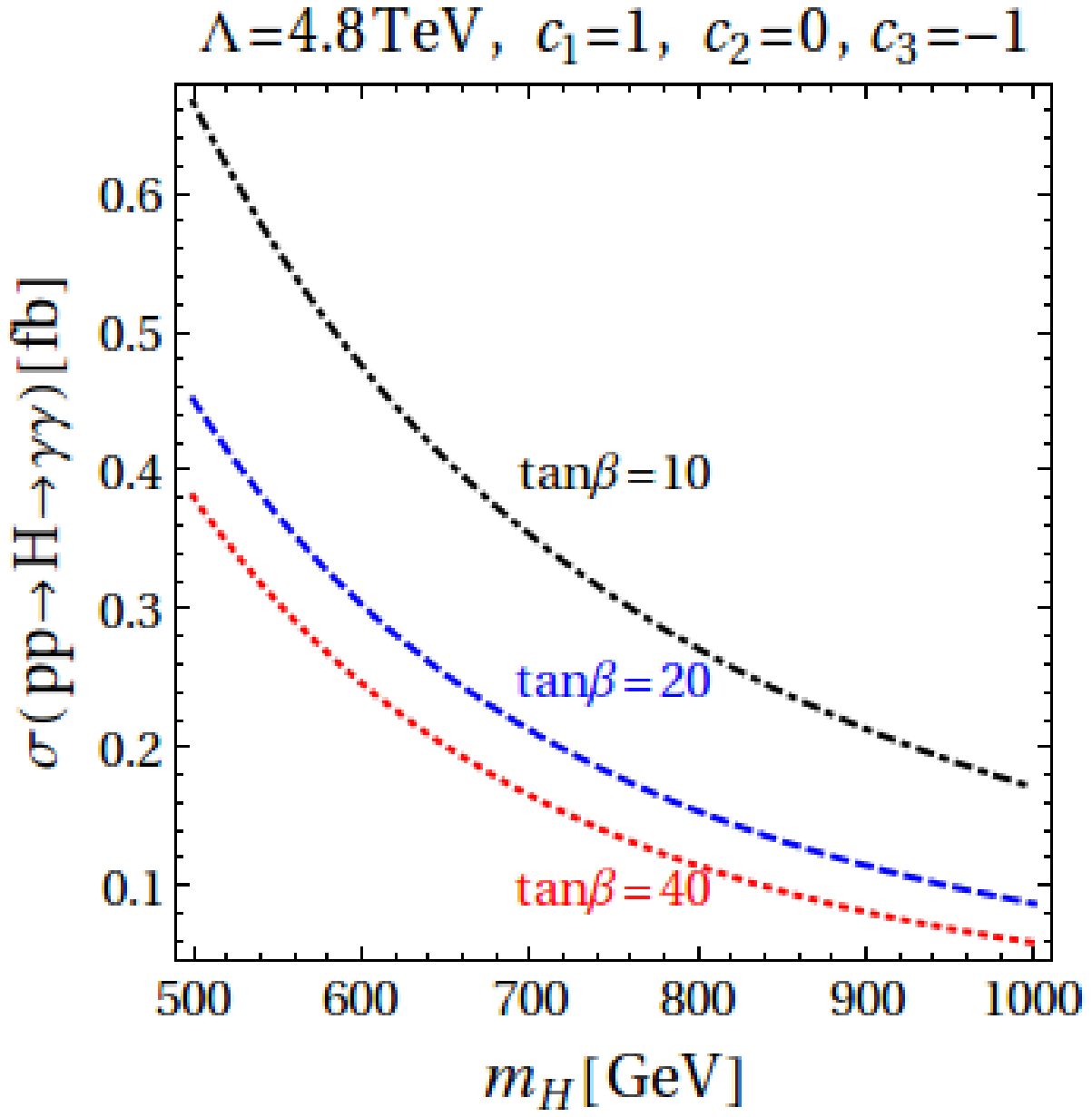} 
 \hspace{0.08cm}
 \includegraphics[width=6.4cm,height=5.4cm]{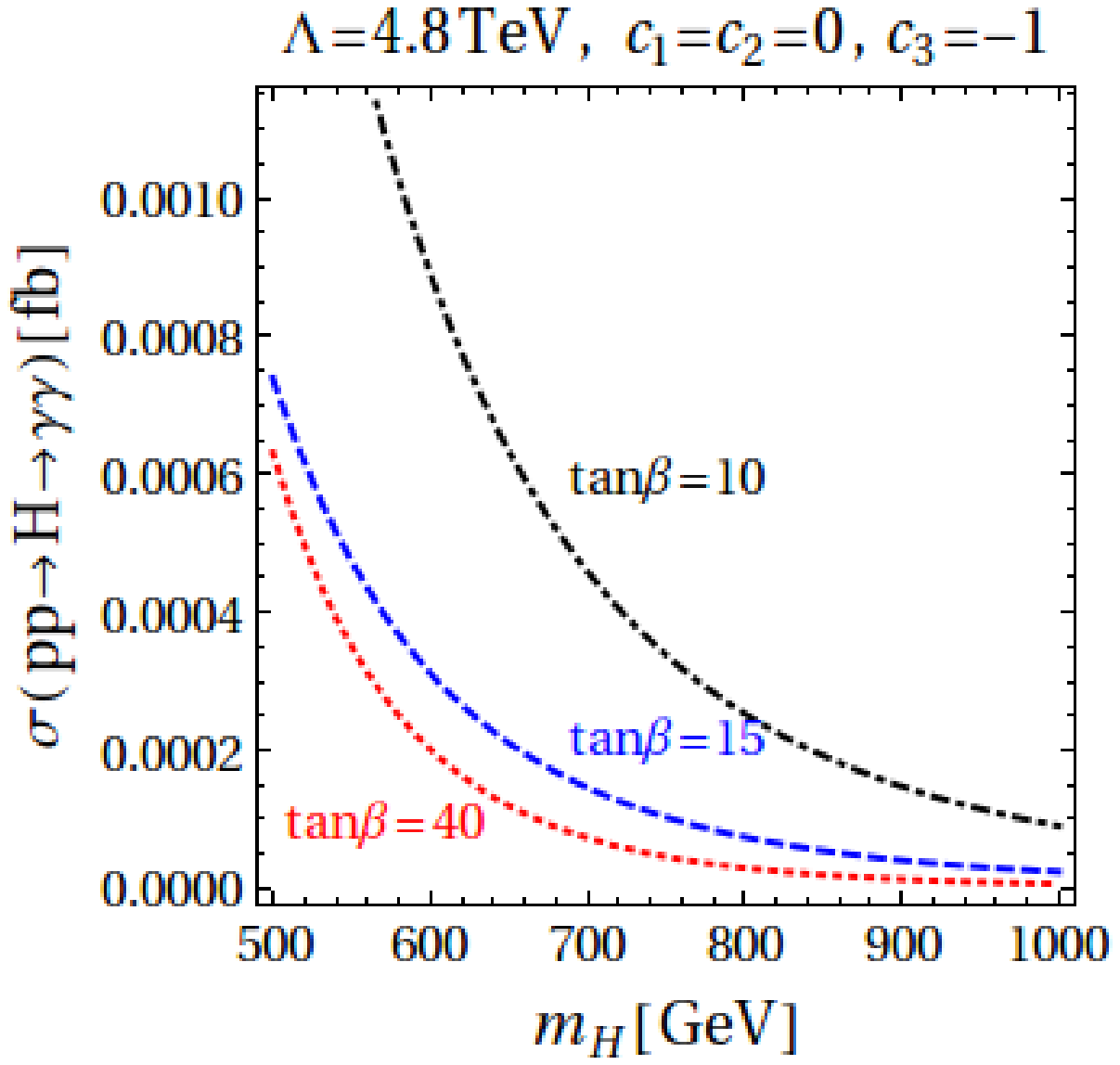} 
\end{center}
\caption{\small The diphoton production cross section for  fixed $\Lambda$ as a
function of the mass $m_X$ ($X=H$) for different $\tan\beta$ and coefficients $c_{1,2,3}$ of the
effective operators. Similar dependence (nearly identical) exists for $X=A$.  
The largest cross section for a given $\Lambda$ 
is obtained if all $c_{1,2,3}\not =0$ and have  appropriate relative signs (shown).
 Of individual contributions
for the same effective scale, the largest correction to $\sigma$ is from 
$\cL_3$, then $\cL_1$.
}
\label{crosss}
\end{figure}

\begin{figure}[!h]
   \begin{center}
 \includegraphics[width=6cm,height=5.4cm]{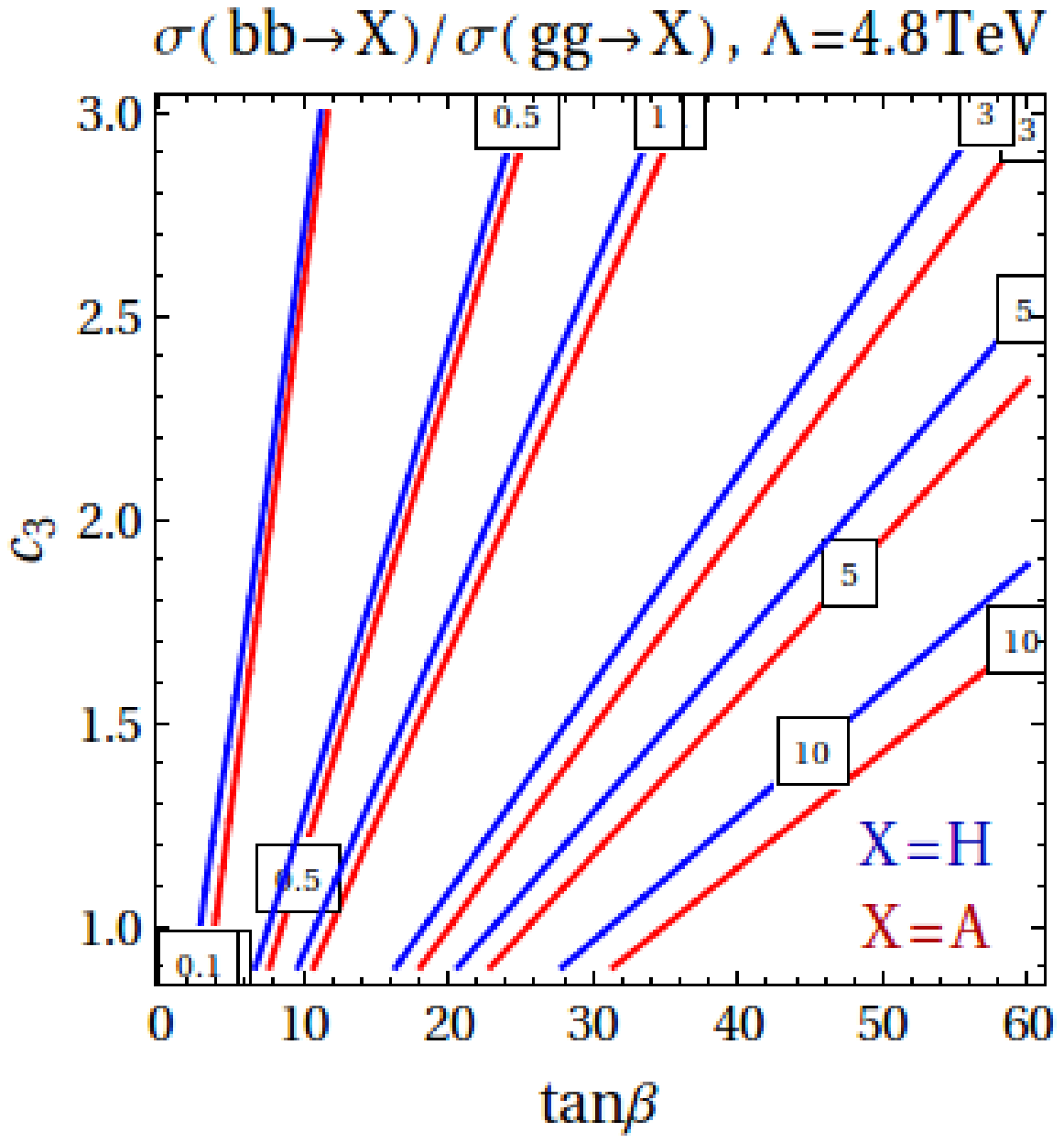} 
 \hspace{0.3cm}
 \includegraphics[width=6cm,height=5.4cm]{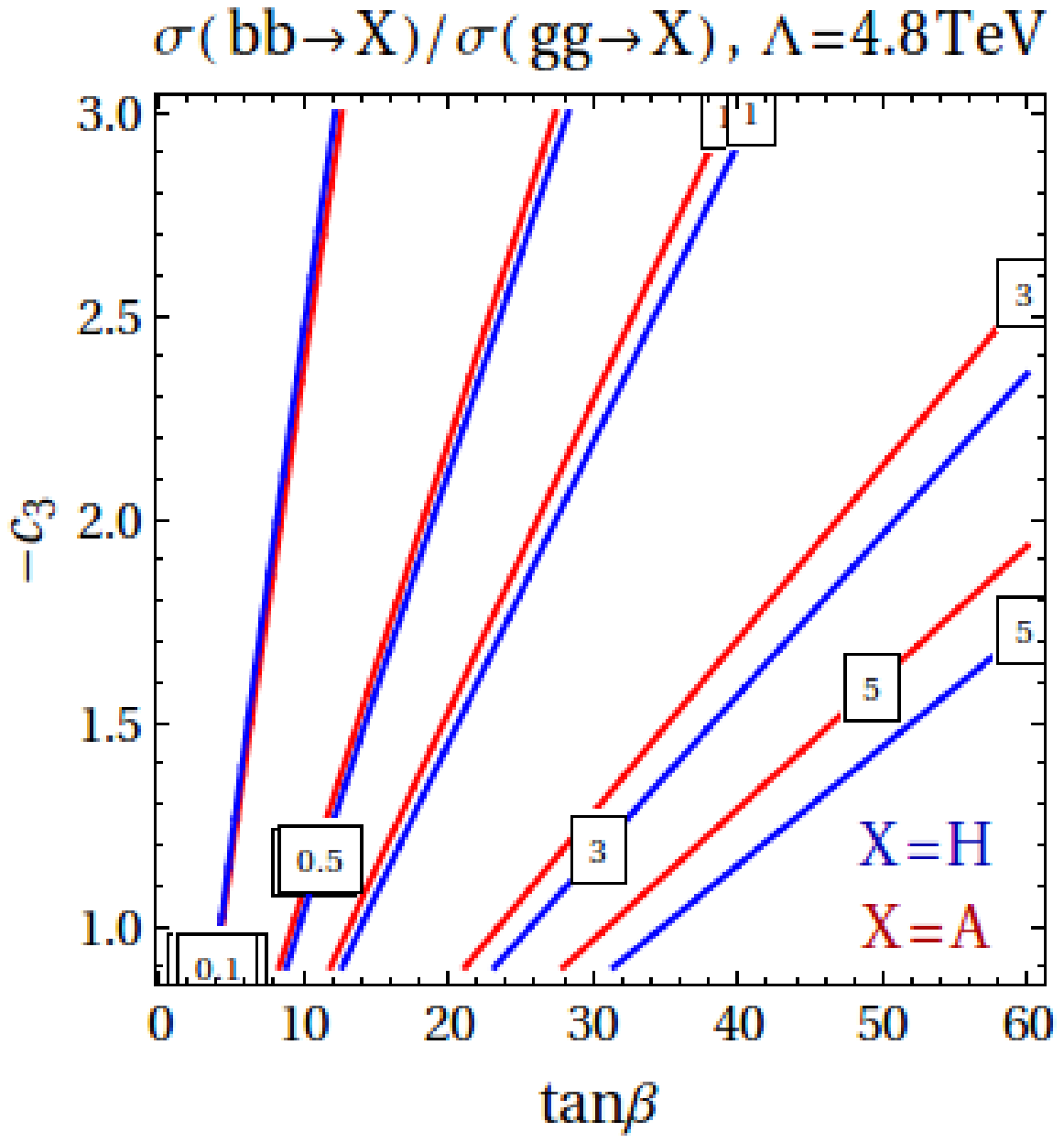} 
    \end{center}
\vspace{0.0cm}
   \caption{
\small
 Ratios of  production cross sections:  
 $\sigma(b{\bar b}\rightarrow X)$/$\sigma(gg\rightarrow X)$ for $X=H$ in blue and $X=A$ in red,
 for $c_3\!>\!0$ (left plot) and $c_3\!<\!0$ (right plot). 
Depending on $\tan\beta$ and $\vert c_3\vert$ either $gg$ or  $b\bar b$  production 
mechanism may dominate or they have comparable  cross sections.
 Here we chose $m_{H,A}=750$ GeV, for illustration.}
  \label{fig5}
 \end{figure}

Using the information in figure~\ref{figc} for the coefficients $C_{gg}$ and $C_{b\bar b}$ 
 one can compute the diphoton production cross section for different
values of the resonance mass. This dependence is shown in the plots of 
figure~\ref{crosss}, for a fixed scale
$\Lambda=4.2$ and $4.8$ TeV
 of the effective operator and different $\tan\beta$ and $c_{1,2,3}$
\footnote{We keep $c_j$ close to unity 
 (while freely adjusting $\Lambda$), otherwise 
 the effective scale of new physics (operator $\cL_j$) is 
 changed to  $\Lambda/\sqrt{\vert c_j\vert}$.}.
Both production channels $b\bar b$ and $gg$ of $H,A$ contribute, see figure~\ref{fig5}.
In some cases, the cross section can be large, $\sigma \sim$ few fb, well 
above its value in the MSSM alone and
this can conflict with the latest data \cite{ncms}. To avoid this situation, 
as seen in figure~\ref{crosss}, a larger $m_X$ and/or larger $\Lambda$ and/or large
 $\tan\beta$ may be required, correlated as shown. 
If the value of $\sigma$ is  known from experiments and assuming $X=H,A$, 
 these plots together with  constraints on SM-like Higgs physics
 can be used to set stronger  bounds on the correlation of $\Lambda$
with  $\tan\beta$ and $m_{H,A}$. We shall do this shortly  for   $m_{H,A}$ in
the range $0.5-1$ TeV.

\begin{figure}[t!]
  \begin{center}
    \includegraphics[width=6cm,height=5.4cm]{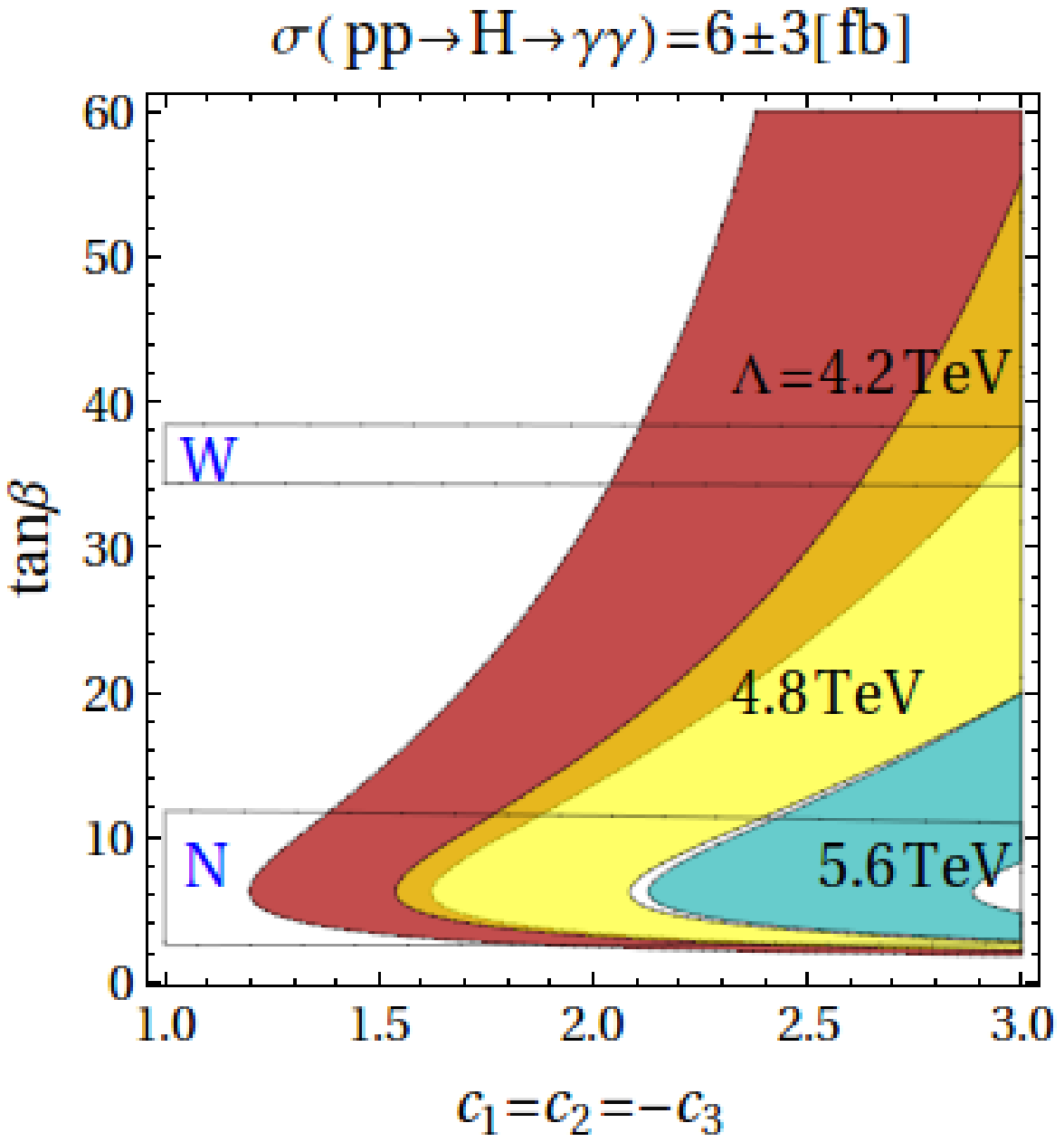} 
 \hspace{0.3cm}
 \includegraphics[width=6cm,height=5.4cm]{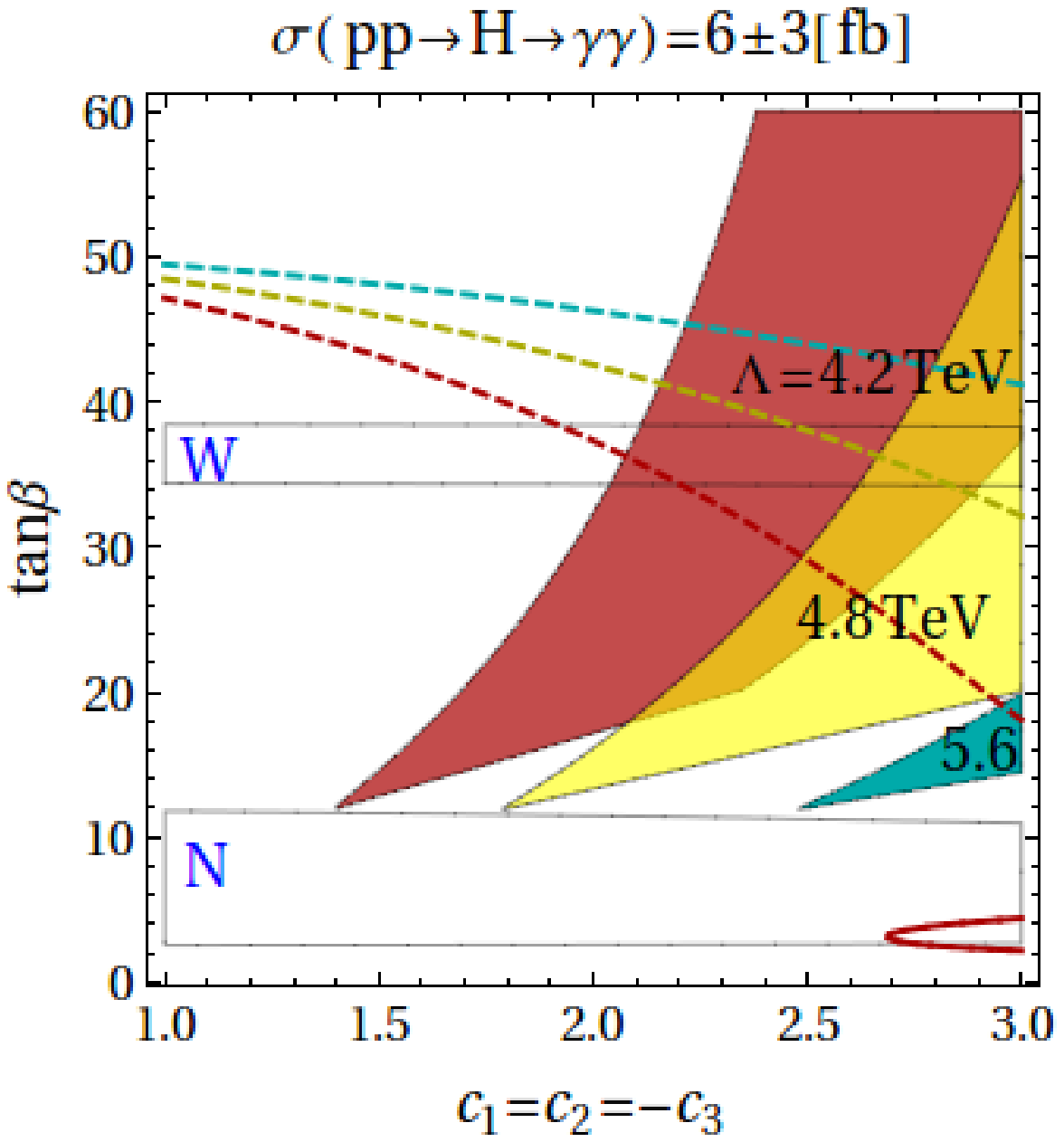} 
  \\[2pt]
 \includegraphics[width=6cm,height=5.4cm]{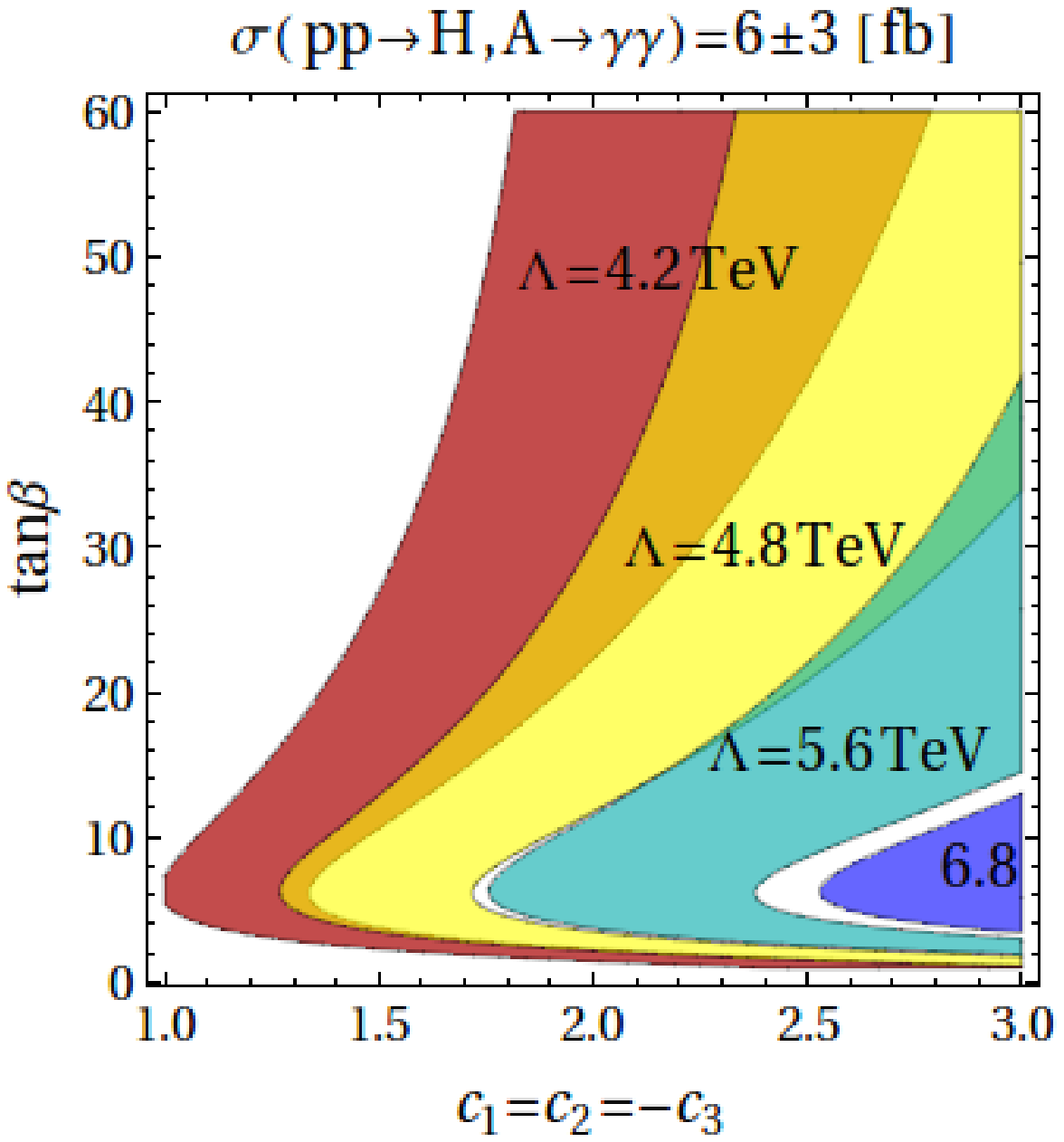} 
\hspace{0.3cm}
 \includegraphics[width=6cm,height=5.4cm]{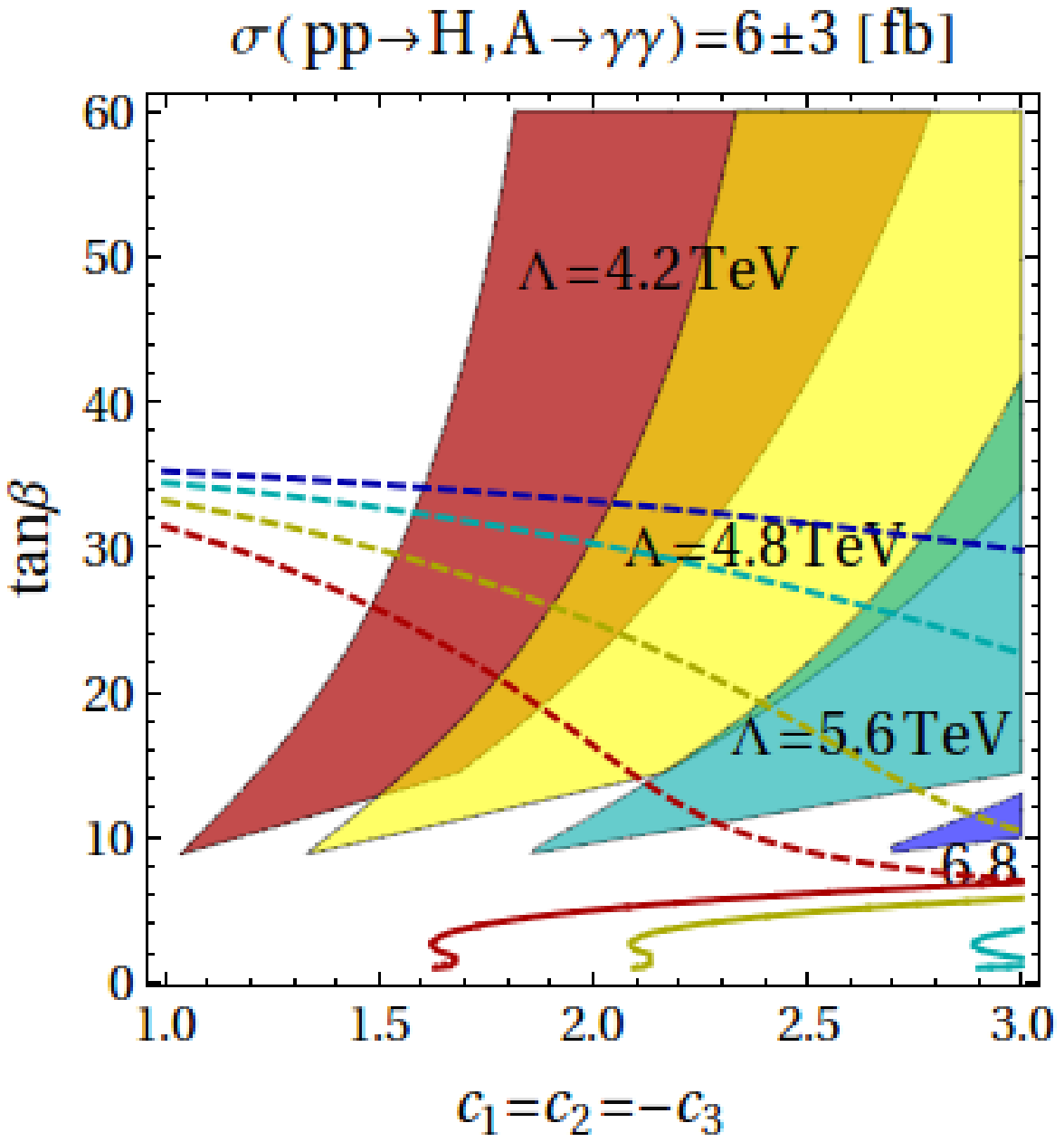} 
 \end{center}
 \caption{
\small
Parameter space for the diphoton cross section $\sigma$, 
mediated by  $H$ (top) and by both $A$, $H$, 
mass degenerate,  at $m_{H,A}=750$ GeV (bottom). 
Nearly identical plots to $H$ exist for $A$ alone.
Each coloured area has $\sigma=6\pm 3\,{\rm fb}$ 
with a fixed  $\Lambda$.
Areas of overlapping colours (e.g. red and yellow, shown in orange)
 correspond  to two  different $\Lambda$, ($3\leq \sigma\leq 9$).
Region  ``N'' corresponds to a 
narrow resonance $\Gamma_{H,A}\leq 5$ GeV;
region   ``W'' has a wide width of $40\leq \Gamma_{H,A}\!\leq\! 50$ GeV.
For  $12\!\leq\! \tan\beta\!\leq\! 36 $, $\Gamma_{H,A}$ has intermediate values
(fig.\ref{fig2}). 
Region ``W'' is excluded by the
constraint $\cR_{b\bar b} \equiv \Gamma_{X\ra b\bar b}/\Gamma_{X\ra \gamma\gamma}
\leq 500 (r/5)$, $X\!=\!A,H$. 
The plots in the right column have  CMS  bounds on $\kappa_\gamma$ and $\kappa_g$
applied which excluded  the low $\tan\beta$ regions. 
A dotted (continuous) curve in a colour rules out the area (if present) 
in the same colour situated above (below) that curve,
due to $b\bar b$ ($t\bar t$) searches, respectively \cite{GG}. The
corresponding region of this  ``resonance'',  for fixed $\Lambda$, 
is the area in a given colour below the 
 dotted curve {\it in the  same colour}, in the right column of plots.
}
\label{fig3}
\end{figure}

\subsection{The  dependence of $\sigma$ on  $m_{H,A}$ and the missing 750 GeV ``resonance''}

As seen in the previous section,
$\cL_{1,2,3}$ may provide a diphoton production cross section that
is as large as few fb, as initially reported by ATLAS/CMS \cite{750}, at  $m_X=750$ GeV,
($X=H,A$), now ruled out by recent additional data \cite{ncms}. 
In the following  we first detail the exclusion  limits on the scale  $\Lambda$, 
correlated with  $\tan\beta$, from the absence of this resonance.
We then consider the  general case of varying $0.5$ TeV$\leq  m_{H,A}\leq 1$ TeV 
and explore the dependence of  $\sigma$ on $m_{H,A}$,  $\Lambda$ and 
$\tan\beta$, under the experimental constraints from the SM-like higgs couplings
$hgg$ and $h\gamma\gamma$
and from  $b\bar b$ and $t \bar t$ searches. Both production channels of
$X=H,A$ are included and either of these may dominate (figure~\ref{fig5}).

Figure~\ref{fig3} shows  the parameter space  giving the
initially found $\sigma(pp\ra H,A\ra \gamma\gamma)=6\pm 3$\,fb at $m_X=750$ GeV,
 mediated by $H$ or $A$ or both (mass degenerate case\footnote{In the
decoupling limit we use, valid for $m_A=750$ GeV, the mass splitting $\delta m$
between $A$ and $H$  can be neglected
$m_H^2=m_A^2+m_Z^2 \sin^2[2\beta]$, so $\delta m\leq 2$ GeV  for $\tan\beta>3$
and decreases further at larger $\tan\beta$.}).
The allowed parameter space is  similar for $A$ and $H$.
In this figure the relative signs of  $\vert c_j\vert \sim 1$ were chosen to
maximise the diphoton production for given $\Lambda$. 
Note that  the effective cutoff of an operator is ultimately
 $\Lambda/\sqrt{\vert c_j\vert}$.

{\bf  Narrow resonance:\,\,} 
For low $2\leq \tan\beta\leq 12$  (figures~\ref{fig2}, \ref{fig3}) 
one has a narrow width, $\Gamma_{H,A}\leq 5$ GeV. 
For $\tan\beta\leq 8$,  the $gg$ production channel of $H, A$ dominates;
 for $8\leq \tan\beta\leq 12$ the  $b\bar b$ channel is also relevant (figure~\ref{fig5}).

Let us see the effect of the constraints from the SM-like higgs ($h$) rates.
In figure~\ref{fig3} the low $\tan\beta$ region
contributes to $h\ra\gamma\gamma$ (photons) and $h\ra gg$ (gluons) and can  even
enhance (reduce)  the rate of $h\ra \gamma\gamma$ beyond the SM value for negative
 (positive) $c_{1,2}$,  respectively \cite{Berg}\footnote{
This is due to coefficients $\hat c_{\gamma\gamma}$ or $c_{\gamma\gamma}$ which
contribute  at low $\tan\beta$, see eqs.(\ref{R12}), (\ref{dd}) 
for $\alpha\ra \beta-\pi/2$.}.
Define by $\kappa_\gamma$ and $\kappa_g$ 
the scaling coefficients of the amplitude of the SM-like higgs couplings 
to  $\gamma\gamma$  and  $gg$; then  one has \cite{RPP} (see also \cite{kg,E}):
\bea
\label{kappa}
{\rm CMS,} \,\, 68\% {\rm CL:} & \qquad&  \kappa_\gamma=0.965\pm 0.175,  \qquad \kappa_g=0.835\pm 0.105\\
{\rm ATLAS,} \,\,68\% {\rm CL:} & \qquad& \kappa_\gamma=1.2\pm 0.15,\quad\,\,\,\qquad \kappa_g=1.04\pm 0.14
\eea
We used the CMS constraint in fig.~\ref{fig3} 
 at $95\%$ CL, with  $\kappa_j^2=\Gamma_{h\ra jj}/\Gamma_{h\ra jj}^{SM}$, $j=\gamma, g$.
As a result, $2\leq \tan\beta\leq 10$ or so is in conflict with these constraints from
$h$ decays and this region, largely overlapping with our  narrow width regime,
is ruled out. Further, $t\bar t$ searches  also rule out some parameter space
close to  $1\leq \tan\beta<8$ but  the bound found is in general weaker than the above 
bounds from $h$ signals\footnote{The bound used for $t\bar t$ searches is 
$\sigma(pp \ra X \ra  t\bar t)<2250$ fb (13 TeV), see Table 1 in \cite{GG}.}.
As a result, the  parametric region in  figure~\ref{fig3} 
corresponding to this narrow ``resonance'' 
is a small region at the tip of each coloured area of fixed $\Lambda$
 with  $\tan\beta\!\approx\! 10-12$.

{\bf Broad resonance:\,\,\,}  
The region $34\leq \tan\beta\leq 38$ ($40\leq \Gamma_{H,A}\leq 50$ GeV, fig.\ref{fig2})
marked  as ``W'' in figs.~\ref{fig3}, where  the $b\bar b$ production mechanism 
dominates (if $\vert c_3\vert \approx 1-2$, fig.\ref{fig5}),
 is ruled out by constraints such as those in  Table~\ref{table2}, of 
which $\cR_{b\bar b}<500$ is the strongest.
Further, $b \bar b$ searches with a cross section bound $\leq 5$ pb at 13 TeV 
(Table 1 in \cite{GG}) are also marked in figs.\ref{fig3}, 
with a  dotted curve in a given colour  that rules out any area
 {\it in the same colour} situated {\it above} that curve.
This leaves a parametric area bordered by $\tan\beta\leq 25$ with $\Gamma_X\leq 25$ GeV 
($\tan\beta\leq 18$, $\Gamma_X\leq 12$ GeV) for $\sigma\approx 3$ fb
($\sigma\approx 9$ fb) respectively, for  mass degenerate $A$ and $H$ 
and  $\Lambda/\sqrt{\vert c_j\vert}$ fixed.

With this resonance now ruled out, one must then exclude its parametric region 
bordered by  $10\leq \tan\beta\leq 25$ ($10\leq \tan\beta\leq 18$)
for $\sigma\approx 3$ fb ($\sigma\approx 9$ fb), respectively and 
 demand the effective scale be larger than  $\Lambda/\sqrt{\vert c_j\vert}\approx 4.2$ TeV. 
We checked that similar bounds apply for mildly different values of $c_{1,2,3}$ and 
from unity.
These bounds  are relevant provided that all  $\cL_{1,2,3}$  in eq.(\ref{dim6})  contribute. 
Since  $\cL_3$ is the dominant contribution,  if  $c_3=0$ then  one has a much smaller 
diphoton cross section.
If  $c_2=0$  (or $c_1=0$) and $c_3\not=0$  i.e. only $\cL_{1,3}$ ($\cL_{2,3}$) are present,
total $\sigma$ is again reduced; one may still reach  $\sigma\sim 3$ fb
 by compensating with an increase  of the remaining coefficients,  but then
$\Lambda/\sqrt{\vert c_j\vert}$ may become  too low for a 
reliable effective expansion.

\vspace{1cm}

\begin{figure}[t!]
  \begin{center} 
\includegraphics[width=6cm,height=5.4cm]{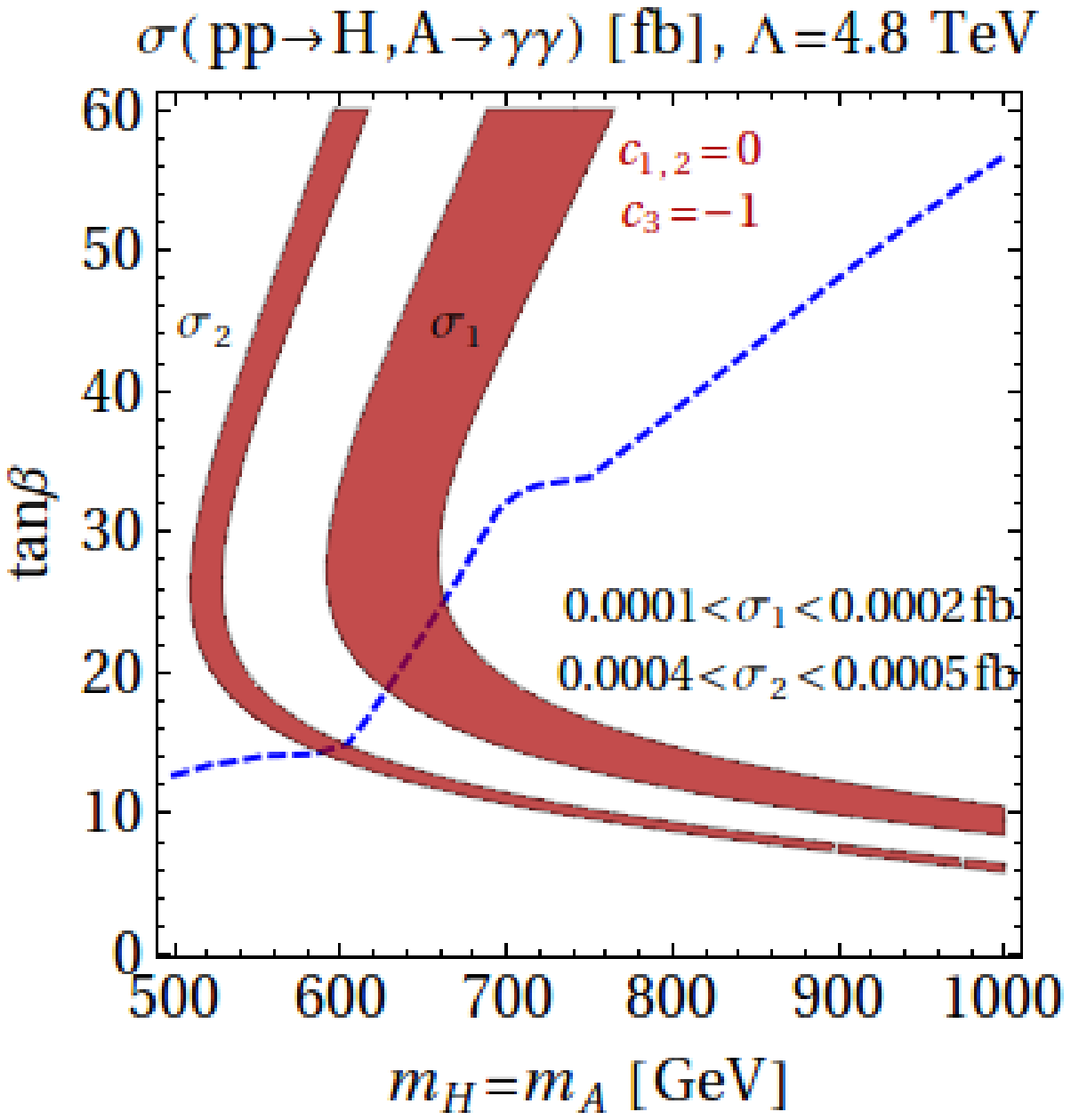} 
\hspace{0.3cm} 
\includegraphics[width=6cm,height=5.4cm]{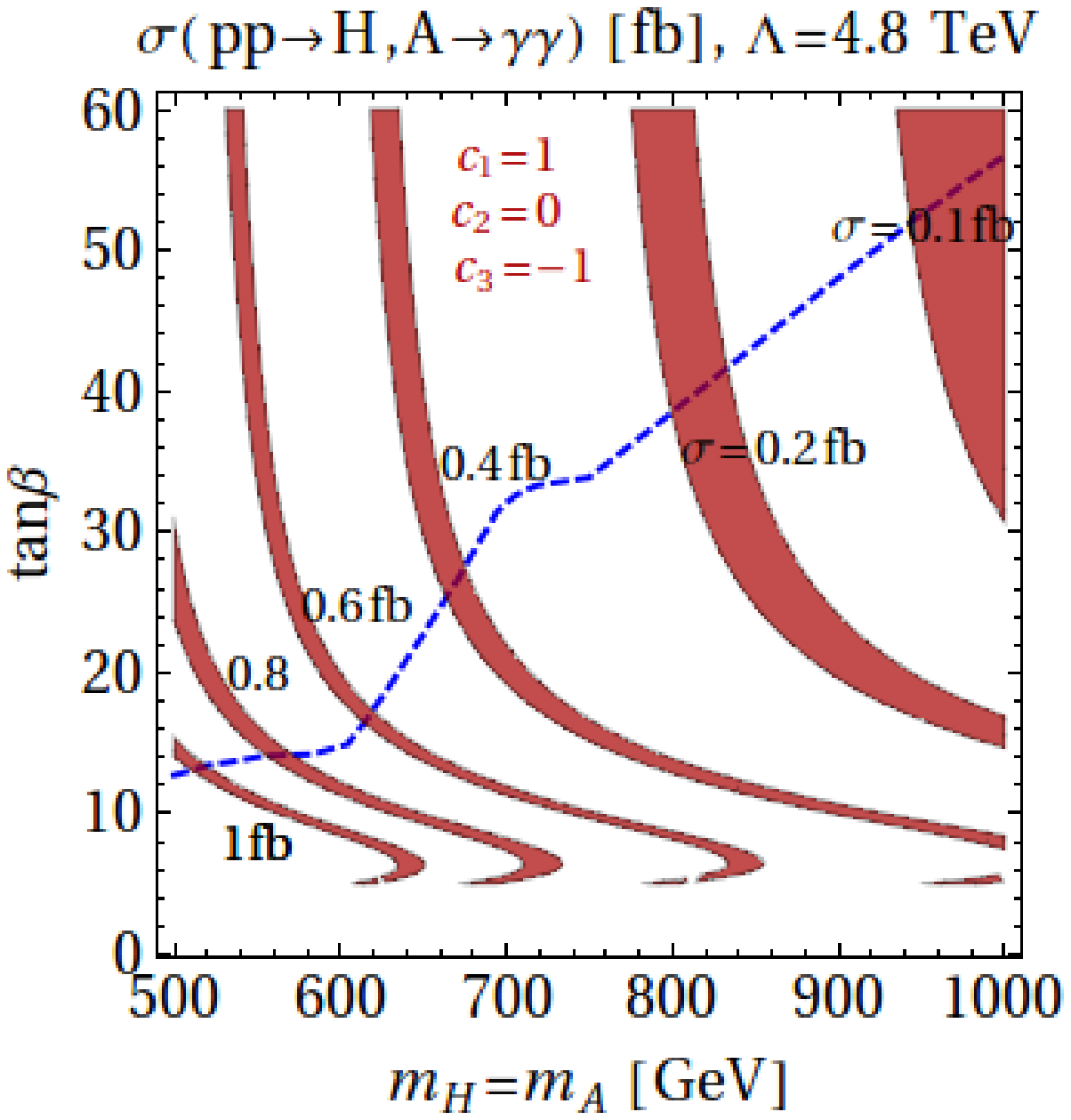} 
 \end{center}
 \caption{
\small
Parameter space for different values of the diphoton cross section  $\sigma$, 
mediated by  both $A$, $H$ (mass degenerate) for a  varying mass $m_{H,A}$
with $\Lambda=4.8$ TeV and  $c_{1,2}=0$, $c_3=-1$ (left)
and $c_1=-c_3=1$, $c_2=0$ (right).
The red regions in  the left plot have $\sigma$  in the range shown
while in the right plot,  their left edge  has $\sigma$ larger by $0.03$ fb
from the values shown.  
The  partonic integral  coefficients dependence on $m_{H,A}$ is included.
The CMS  bounds  on $\kappa_\gamma$ and $\kappa_g$ 
excluded a small low $\tan\beta<6$ regions. 
 The dotted curve (in blue) corresponds to a bound from $b\bar b$  searches (observed values, see
 figure 6 in \cite{bb}). The $t\bar t$ searches bound (observed values, see figure 2 in \cite{tt}) is also
 imposed but for the cases considered here for $c_{1,2,3}$ does not constrain the 
 parameter space.  The allowed parametric region is then the area below the  dotted curve.
}
\label{extra}
\end{figure}

We return now  to  a general case of varying $m_{H,A}$ in the range
 $0.5$\,TeV $\leq\! m_{H,A}\!\leq 1$ TeV.
   Figure~\ref{extra} shows  the dependence of the diphoton cross section
 $\sigma$ on $m_{H,A}$ and $\tan\beta$, under the experimental
constraints from $hgg$ and $h\gamma\gamma$ couplings and $b\bar b$, $t\bar t$ searches.
The cross section  bounds for  the  $b\bar b$ and $t\bar t$ searches depend on $m_{H,A}$; 
 we used the observed values ($95\%$ CL) for  $b\bar b$ searches of  figure~6 in  \cite{bb} and for $t\bar t$ 
 searches of  figure~2 in \cite{tt}, for the range of $m_X$ considered in our figure~\ref{extra}. These
 values were scaled to $\sqrt s=13$ TeV.
The dependence of the parton coefficients on $m_{H,A}$ is also included
(see  figure~\ref{figc}).

Large   values of diphoton cross section $\sigma\sim 0.1 - 1$ fb (well above the  MSSM value) are
obtained when both $\cL_3$ and $\cL_1$ are present, for  $\Lambda=4.8$ TeV (right plot in 
fig.\ref{extra}).  For $\vert c_3\vert$ only mildly 
different from unity $\vert c_3\vert \sim 1.3$ or 
if also $c_2\not= 0$, then  $\sigma$ increases further from the values shown.
Unlike for the $750$ GeV ``resonance'',  there are now regions of  
low $\tan\beta<10$ with a large diphoton production  such as:
$\sigma\sim 1$~fb for  $m_{H,A}\sim 550 - 650 $ GeV, or $\sigma\sim 0.4$ fb for $m_{H,A}\sim 1$ TeV,
that pass all the above  constraints\footnote{The low $\tan\beta$ region may also 
be interesting for the  naturalness issue, see later.}. 
Increasing $\Lambda$ above $\sim 5$ TeV or considering instead only individual operators, 
e.g. dominant $\cL_3$ (left plot in figure~\ref{extra}), can reduce  $\sigma$ significantly. 
This ends our analysis
of the diphoton cross section for $m_{H,A}$ in the range $0.5-1$ TeV.


\begin{table}[!t]
\centering
\begin{tabular}{|cccccc|}
 \hline
$\cR_{ZZ}$  &   $\cR_{Z\gamma}$ &  $\cR_{WW}$  &  $\cR_{t\bar t}$  & $\cR_{b\bar b}$ &  $\cR_{gg}$\\
& & & & &\\[-12pt]
$6 (r/5)$  &   $6 (r/5)$    &   $20 (r/5)$  & $300(r/5)$   & $500(r/5)$   &  $1300 (r/5)$\\
 \hline
\end{tabular}
\caption{\small
Upper bounds on the partial widths 
$\cR_{ab}=\Gamma_{X\ra ab}/\Gamma_{X\ra \gamma\gamma}$, ($X=H, A$),  obtained
 from 8 TeV data scaled to 13 TeV, with
 $r=\sigma_{\rm 13 TeV}/\sigma_{\rm 8 TeV}\approx 5$\, 
and a wide resonance $\Gamma_{H,A}=45$ GeV \cite{GG}. 
They  apply at large  $\tan\beta$.
 $\cR_{b\bar b}$ is the strongest bound.}
\label{table2}
\end{table}

\vspace{2cm}

\section{Microscopic models for $\cL_{1,2,3}$ and higgs mass corrections}
\label{higgs}

Having seen the role  of $\cL_{1,2,3}$ on the diphoton
 cross section, we now explain  their possible origin in a renormalizable model.
We also address their effect  on the higgs sector masses.

\subsection{Microscopic origin of effective operator(s) $\cL_{1,2,3}$}

 $\cL_{1,2,3}$  may be generated in the MSSM with additional states with  mass of order $\Lambda$.
To see this, consider a massive gauge singlet $S$ that couples to 
 the higgs and gauge sector  as in:
\medskip
\bea
\label{gen}
\delta L\!\! =\!\!
\int \! d^4\theta\, S^\dagger S+\Big\{\!\!
\int\! d^2\theta \Big[ \mu H_1.H_2\! +\!\lambda \,S\,H_1.H_2\! + \frac{1}{2}\, M_1 S^2
+ f(S)\Tr(\cW^\alpha W_\alpha)\Big]\!+\!\textsf{h.c}\Big\}
\eea

\medskip\noindent
 $f(S)=S/M_2$ is  a gauge kinetic function of a SM subgroup and $M_2$ some high mass scale. 
To generate all  $\cL_{1,2,3}$ the coupling to the gauge sector is extended to 
$SU(3)\times SU(2)_L \times U(1)_Y$.
We  integrate  out the superfield $S$ via its eqs of motion and find 
 after some algebra and consistent truncation of higher orders 
\bea\label{W}
\delta L\!\!\!&=&\!\!\!\!
\int d^4\theta\,
\Big[
 2\,\Big\vert \frac{\lambda}{M_1}\Big\vert^2\, \vert  H_1.H_2\vert^2
\Big]
\nonumber\\
&&\!\!\!\!\!\!\!\!\!\!\!\!\!\!\!\!\!
+ \int\!\! d^2\theta \Big[ \mu\, H_1.H_2-\frac{\lambda^2}{2\,M_1}\,(H_1.H_2)^2
\!-\frac{\lambda}{M_1\,M_2} (H_1.H_2)\, \Tr(W^\alpha W_\alpha)\Big]\!+\!\textsf{h.c}
\!+\!\cO(1/M_1^3)\quad
\eea

\begin{figure}[t]
  \begin{center}
\includegraphics[width=10cm,height=3cm]{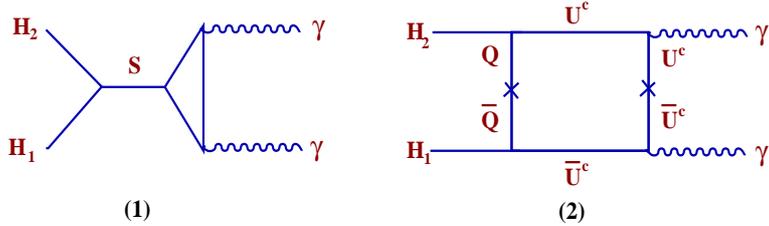}
   \end{center}
\vspace{-0.3cm}
  \caption{\small
Diagrams generating the $d=6$ operator
$(H_1.H_2)\,\Tr\,(W^\alpha W_\alpha)\vert_F$ at one-loop level. 
Diagram (1) corresponds to the approach
 in eqs.(\ref{gen}), (\ref{W}),  with a loop of  vector-like states
of mass $\propto M_2$ that generate  $(S/M_2)\Tr(W^\alpha W_\alpha)\vert_F$ of eq.(\ref{gen}),
 while (classical) integration of $S$   generates the needed $d=6$ operator. 
Diagram (2) (plus another one  as (2) but  with
$Q\leftrightarrow U^c$, $\overline Q\leftrightarrow \overline U^c$)  
show how to generate the $d=6$  operator in one stage, without a massive singlet 
$S$. A large number of vector-like states  can compensate the loop suppression.}
  \label{fig7}
\end{figure}

\noindent
In the rhs of the above equation there are two more terms:
 $-1/(2M_1 M_2^2) (\Tr W^2))^2\vert_F$ and also  $2\lambda/(M_1^2 M_2)(H_1.H_2)^\dagger \Tr W^2\vert_D$;
since  we choose  $M_2\geq M_1$, they are sub-leading, $O(1/M_1^3)$, and can be ignored.
The last line in eq.(\ref{W}) shows our operators of $d=6$ and $d=5$ 
generated simultaneously when integrating $S$.
However eq.(\ref{gen}) does not yet provide a UV complete, 
renormalizable setup, since it still contains a  $d=5$ effective  operator:
 $(S/M_2)\,\Tr\, (W^\alpha W_\alpha)\vert_F$.
One possibility is that this operator is generated  if
 $S$ has additional, renormalizable couplings to massive vector-like 
states under the SM gauge group, of mass $\cO(M_2)$, as shown in  diagram  (1)  
of figure~\ref{fig7}.
Integrating out the vector-like states then 
 generates this remaining operator\footnote{ $(S/M_2)\,\Tr\, W^\alpha W_\alpha\vert_F$  is a
 moduli-dependent gauge   kinetic term,   generic in supergravity or string theory.}. 
We thus have a microscopic  origin of $\cL_{1,2,3}$.

Eq.(\ref{W}) also contains a $d=6$ operator $\vert H_1.H_2\vert^2_D$
 which brings  a negative correction to the SM-like Higgs mass
$\delta m_h^2=-4v^2 \vert\lambda\vert^2 \mu^2/\Lambda^2+\cO(1/\tan^2\beta)$ 
\cite{d=6} (for $M_1=\Lambda$); this correction is less relevant
(being sub-leading to that of eq.(\ref{dim5}), see later). 
Finally, taking $M_1\sim M_2\sim \Lambda$ and comparing eq.(\ref{W}) to (\ref{dim6}), we 
identify $\lambda= c_j/2$ and $c_0=-\lambda^2/2=-c_j^2/8$. 

Another way to generate $(H_1.H_2)\,\Tr(W^\alpha W_\alpha)_F$ is
at one-loop,  without a massive singlet.
One considers only copies of massive vector-like states 
as  in diagram (2) of fig.\ref{fig7}.

 To conclude,  a heavy diphoton resonance ($X=H,A$) of large cross section 
is present  if  SM-charged, massive vector-like states
(and possibly a singlet)  are present beyond MSSM;  after decoupling,
they generate $\cL_{1,2,3}$. Other ways to generate the $d=6$ operator(s) may exist.
 The vector-like states have a significant impact on the gauge 
 couplings unification  at one-loop, unless they are complete  
 $SU(5)$ multiplets \cite{unif}.

\subsection{Implications for Higgs sector masses}

Unlike the ``gluon'' operator $\cL_3$, the  ``electroweak''  operators $\cL_{1,2}$ 
of eq.(\ref{dim6}) also impact on the higgs masses
$m_{h,H}^2=M_{h,H}^2+\Delta m_{h,H}^2$.  Here $M_{h,H}$ denote the MSSM  
value. We find\footnote{using the first ref in \cite{d=6} and adding a one-loop effect, too (top Yukawa).}
\bea\label{deltam}
\Delta m_{h,H}^2= \frac{8 \rho}{\Lambda^2}
\Big[ g_1^2c_1 +  g_2^2c_2 \Big] 
+\cO\!\Big[\frac{1}{ \Lambda^3}\!\Big] \,
\eea
where  
\bea
\rho=\frac{v^4}{32}\,\sin 2\beta\,\Big[1\pm \frac{1}{4\,\sqrt w}
\big[ 8 m_A^2-(4+3\delta) m_Z^2+6 \delta m_Z^2 \cos 2\beta
+3 (4 m_A^2-\delta m_Z^2)\cos 4\beta\big]\Big]
\eea

\medskip\noindent
and $w=[(m_A^2-m_Z^2)\,\cos 2\beta+\delta\,m_Z^2\,\sin^2\beta]^2
+\sin^2 2\beta \,(m_A^2+m_Z^2)^2$; the upper (lower) signs correspond to $h$ ($H$)
and $\delta$ is shown in Appendix~B.
 These  corrections bring 
a  modest increase of the SM-like higgs mass  $\Delta m_h\sim  1 \textsf{GeV}$ for  
$c_{1,2}=\cO(1)$, with a largest value   for small $\tan\beta$,
with little dependence on $m_A$. An even smaller correction is found for $m_H$.
The mass of the  CP-odd Higgs boson is also modified, 
see eq.(\ref{mA}). These corrections have little 
impact on the previous diphoton analysis.

As we saw in the previous sub-section,  a  leading $d=5$  operator  
\begin{eqnarray} 
\cL_0& = & \frac{c_0}{
\Lambda}\int d^{2}\theta \,
\,(H_1 . H_2)^{2} + \mbox{h.c.}
\label{dim5}
\eea
may also be generated from the UV complete (renormalizable)  model, 
without direct contribution to  the diphoton cross section. 
Its correction to the higgs mass is \cite{ft,d=6}
\bea\label{deltam2}
\Delta m_{h,H}^2= 
 \left[2\mu\,\frac{c_0}{\Lambda 
}\right] \rho_1  + \left[2\mu \frac{c_0}{\Lambda 
}\right]^2 \rho_2 
+\cO\!\Big[\frac{1}{\Lambda^3}\!\Big] \,
\eea
where $\rho_{1,2}$ are shown in eq.(\ref{md5}).
With $c_0\,\mu>0$, 
a  numerical analysis shows  a significant increase of $m_h$  for small $\tan\beta<10$,   
by as much as $\approx 10$ GeV \cite{ft}. This increase
can  reduce the amount of EW scale fine tuning  by a significant factor \cite{ft,Cassel:2011zd}
relative to its MSSM value at low $\tan\beta$ region  (which 
is an otherwise very fine tuned MSSM region).

\section{Conclusions}\label{conc}

Current  searches for ``new physics'' at the LHC bring  increasingly strong constraints 
on the MSSM-like models. Their parameter space becomes 
smaller, with negative implications for their naturalness. 
However, simple extensions of their  minimal higgs sector, parametrised 
 by effective (supersymmetric) operators,  relax the parameter space
or even improve naturalness. 
We studied the constraints  on  such operators 
that can enhance dramatically  the couplings of the higgs sector to SM gauge bosons
and thus the heavy diphoton production.

Minimal  models (MSSM) have a small  diphoton cross  section at large $m_{H,A}$, 
 unless one is  fine-tuning the parameters.  We identified 
 leading operators of dimension $d\!=\!6$ in the higgs sector
$\cL_j\sim c_j/\Lambda^2\, (H_1.H_2) \Tr (W^\alpha W_\alpha)_j\vert_F$, $j\!=\!1,2,3$, $c_j=\cO(1)$,
that enhance  the couplings to SM gauge bosons, with $\cL_3$ having the dominant effects.
For  $m_{H,A}$ in the range $0.5$ TeV$\leq m_{H,A}\leq 1$ TeV,  the
 combination $\cL_1+\cL_3$ 
 can lead to a large  diphoton production $\sigma\sim 0.1 - 1$ fb,
  well above the MSSM value. 
The analysis included both $gg$ and $b\bar b$ production mechanisms (of 
 $X\!=\! H,A$) and either of these may dominate.
 We  examined the correlation between the diphoton  cross section $\sigma$
and  the values of $m_X$,  $\Lambda$ and $\tan\beta$, 
under the  experimental constraints from  SM-like higgs couplings $hgg$ 
and $h\gamma\gamma$ (due to mixing) and from $b\bar b$ and $t\bar t$ searches.
These give $\Lambda/\sqrt{\vert c_j\vert}> 4.8 $  TeV  where the effective approach
can still be trusted,  for $m_{H,A}$ between $0.5-1$ TeV. 

Regarding the initially claimed
 resonance at $m_X\!=\!750$ GeV  with even larger  $\sigma$ (few fb), this could be
reached if all $\cL_{1,2,3}$  contribute. Recent data ruled out this resonance, 
then  not all $\cL_{1,2,3}$ are simultaneously present or the scale
 $\Lambda/\sqrt{\vert c_j\vert}$ is  larger than $4 - 5 $ TeV.

We  showed how to generate the $d=6$  effective operator(s)
 from a UV complete (renormalizable) theory. This is possible by integrating
out additional massive  SM vector-like states beyond the MSSM spectrum,
 and eventually a  massive singlet too, of mass $\cO(\Lambda)$. 
An additional $d=5$  operator in the higgs sector may also be generated at the same time,
that does not affect directly  the diphoton production, but may improve naturalness.

\bigskip\medskip
\begin{center}
-----------------------------
\end{center}

\newpage
\section*{Appendix:}

\def\theequation{A-\arabic{equation}}
\def\thesubsection{A}
\setcounter{equation}{0}
 \label{appendixA}
\subsection{Loop functions and couplings}

In this section we present the expressions of the coefficients
$a_{\bf ..}^{loop}$, $b_{..}^{loop}$, $c_{..}^{loop}$, used in the text (section~\ref{eff}).
To compute them and to fix the notation, 
 we need the couplings of  MSSM fields $h, H$, $A$  to fermions and gauge bosons. 
These are, in a standard  notation
\medskip
\bea
- \Delta {\cal L} &=& 
  k_t \, \frac{m_t}{v} \,h \overline t\, t 
+ k_b \, \frac{m_b}{v} \,h \overline b\, b 
+  \frac{2 m_W^2}{v}\,(k_w h +\tilde k_w H) W^{+\mu} W^{-}_{\mu} \nonumber 
\\
&&
+ \tilde k_t\,\frac{m_t}{v}\, H\, \overline t\, t
+ \tilde k_b\,\frac{m_b}{v}\, H\, \overline b\, b
+i\bar  k_t\,\frac{m_t}{v}\,A\,  \overline t\,\gamma^5 t
+i\bar k_b\,\frac{m_b}{v}\,A\,  \overline b\,\gamma^5 b
\eea
where
\bea
k_t&=& \frac{\cos\alpha}{\sin\beta},
\quad
k_b=-\frac{\sin\alpha}{\cos\beta},
\quad
k_w=\sin(\beta-\alpha)\nonumber\\[4pt]
\tilde k_t & =& \frac{\sin\alpha}{\sin\beta},
\quad\,
\tilde k_b=\frac{\cos\alpha}{\cos\beta},
\qquad
\tilde k_w= \cos(\beta-\alpha),   \nonumber\\[4pt]
\bar k_t&=& \cot\beta,  \qquad  \bar k_b= -\tan\beta.
\eea

\medskip\noindent
Here $\alpha$ is the mixing angle  in the Higgs sector. In this paper 
we work in the decoupling limit ($m_A$ large).  Then  $\alpha\ra \beta-\pi/2$ and 
 $k_{t,b,w}=1$ while $\tilde k_t=-\cot\beta$, $\tilde k_b=\tan\beta$, $\tilde k_w=0$.
Then we find the coefficients of the effective operators in eqs.(\ref{dd}), (\ref{hat}),
 as follows  \cite{Carena}
\medskip
\bea\label{loop1}
a_{gg}^{loop}\!\!\!\! & = &\!\!\!
\bar k_t\,{\bar A}_g^{(t)} + \bar k_b {\bar A}_g^{(b)}\approx 
(0.0771+0.7064 i)\cot\beta-(0.00281+0.00194i)\tan\beta\,
\nonumber\\[6pt]
\!\!\!
a_{\gamma\gamma}^{loop}\!\!\!&=&  \!\!\!\!
\bar k_w\, {\bar A}_\gamma^{(W)}\! +\! \bar k_t\, {\bar A}_\gamma^{(t)}\! +\! \bar k_b\,
 {\bar A}_\gamma^{(b)}\approx\! (0.13702\!+\! 1.256i)\cot\beta\!+\!(0.00125\!-\!0.00086 i)\tan\beta
\nonumber\\[6pt]
a_{\gamma z}^{loop}&=&a_{zz}^{loop}=a_{ww}^{loop}=0.
\nonumber
\eea
and
\bea\label{loop2}
b_{gg}^{loop}\!\!\! & = &\!\!\!
 \tilde k_t\,{\tilde A}_g^{(t)} + \tilde k_b {\tilde A}_g^{(b)}\approx 
-(0.441+1.112i)\cot\beta+(-0.00538+0.00387i) \tan\beta\,
\nonumber\\[6pt]
b_{\gamma\gamma}^{loop}\!\!\!\!&=&\!\!\! 
\tilde k_w\, {\tilde A}_\gamma^{(W)}\! + \tilde k_t\, 
{\tilde A}_\gamma^{(t)}\! + \!\tilde k_b\, {\tilde A}_\gamma^{(b)}\approx \! -(0.783+1.98i)\cot\beta
+(-0.00239\!+0.00172i)\tan\beta
\nonumber\\[6pt]
b_{\gamma z}^{loop} &=& 
\tilde k_t \tilde A^{(t)}_{\gamma z} +\tilde k_b \tilde A^{(b)}_{\gamma z}
\approx (0.12513+0.32821 i) \cot\beta+(0.00103 -0.000043 i) \tan\beta
\nonumber\\[6pt]
b_{ww}^{loop}&=& b_{zz}^{loop}=0.
\eea
and finally 
\bea\label{loop3}
c_{gg}^{loop}\!\!\!\! & = &\!\!\!
k_t\,A_{g}^{(t)} + k_b\, A_{g}^{(b)}\approx 0.970+0.0894i\,
\\[6pt]
c_{\gamma\gamma}^{loop}\!\!\!&=&  \!\!\!
k_w A_\gamma^{(W)}+ k_t \,A_\gamma^{(t)} + k_b \, A_\gamma^{(b)} \approx -6.51+0.0397i\,
\eea

\medskip\noindent
where coefficients $A^{()}_{.}$ are one-loop form factors, presented below.

For $h$ one has the following form factors
\begin{eqnarray}
A_{g}^{(\xi)} & =& \frac{3}{4} A_{1/2}(\tau_\xi), \quad\qquad   \xi=t, b.
\\[5pt]
A_{\gamma}^{(\xi)}&=& N_c Q_\xi^2 A_{1/2}(\tau_\xi), \quad\,\,  \xi=t, b.
\\[5pt]
A_{\gamma}^{(W)} &=& A_{1}(\tau_W)  
\\[5pt]
{A}_{Z\gamma}^{(W)} & = &  \cos\theta_w\,A_1(\tau_W,\lambda_W) 
\\[5pt]
 {A}_{Z\gamma}^{(t)} &=& \frac{N_c Q_t}{\cos^2\theta_w}
\frac{(2 T_3^{(t)}-4Q_t\sin^2\theta_w)}{ \cos\theta_w} A_{1/2}(\tau_t,\lambda_t)
\eea
and for $H$
\bea
\qquad\qquad\qquad\qquad
{\tilde A}_{g}^{(\xi)}& = & \frac{3}{4} A_{1/2}({\tilde \tau}_\xi),\qquad\qquad \xi=t, b. 
\\[5pt]
{\tilde A}_{\gamma}^{(\xi)} &=& N_c Q_\xi^2 A_{1/2}(\tilde \tau_\xi), \qquad \xi=t, b.
\\[5pt]
\tilde A_{\gamma z}^{(\xi)}&=&
\frac{N_c Q_\xi}{\cos\theta_w} (2 T_3^{(\xi)} -4 Q_\xi \sin^2\theta_w)
A_{1/2}(\tilde \tau_\xi,\lambda_\xi),\quad \xi=t, b.\qquad\,\,
\eea
and for $A$:
\bea
\bar A_{g}^{(\xi)} & = & \frac{3}{4} \bar A_{1/2}(\bar\tau_\xi),\qquad \xi=t,b. 
\\
\bar A_{\gamma}^{(\xi)} & =& N_c Q_\xi^2 \bar A_{1/2}(\bar\tau_\xi),\qquad \xi=t,b.
\qquad\qquad
\end{eqnarray} 

\medskip\noindent
where $\tau_i=4m_i^2/m_h^2$,  $\tilde\tau_i=4m_i^2/m_H^2$, $\bar\tau_i=4m_i^2/m_A^2$,
 $N_c=3$, $Q_t=2/3$, and $Q_b=-1/3$.   
  $T_3^{(t)}=1/2$, $T_3^{(b)}=-1/2$ and $\lambda_\xi=4 m_\xi^2/m_Z^2$. Finally
\medskip
 \begin{eqnarray}
A_{1/2}(\tau) & = & 2\tau^{2} \left[\tau^{-1} +(\tau^{-1} -1)f(\tau^{-1})\right]\,  \ ,  
\nonumber
\\
{\bar A}_{1/2}(\tau) &=&\tau f(\tau^{-1}), \nonumber
\\
 A_1(\tau) & = & -\tau^{2} \left[2\tau^{-2} +3\tau^{-1}+3(2\tau^{-1} -1)f(\tau^{-1})\right]\,  \ , 
 \nonumber
 \\
 A_{1/2}(\tau,\lambda) & = & I_1(\tau,\lambda)-I_2(\tau,\lambda)\ , 
\end{eqnarray}
 where
\medskip
 \bea
 I_1(\tau,\lambda)&=&\frac{\tau\lambda}{2(\tau-\lambda)}
 +\frac{\tau^2\lambda^2}{2(\tau-\lambda)^2}
 \left[f(\tau^{-1})-f(\lambda^{-1})\right]+\frac{\tau^2\lambda}{ (\tau-\lambda)^2}
 \left[g(\tau^{-1})-g(\lambda^{-1})\right] \ , 
 \nonumber\\
 I_2(\tau,\lambda)&=& -\frac{\tau\lambda}{2(\tau-\lambda)}
 \left[f(\tau^{-1})-f(\lambda^{-1})\right] \ ,
 \eea
 and
\begin{eqnarray}
f(x)&=&\left\{
\begin{array}{ll}  \displaystyle
\arcsin^2\sqrt{x} & x\leq 1
\\
\displaystyle -\frac{1}{4}\left[ \log\frac{1+\sqrt{1-x^{-1}}}
{1-\sqrt{1-x^{-1}}}-i\pi \right]^2 \hspace{0.5cm} & x>1 \ ,
\end{array} \right.
\eea
\bea
g(x)&=&\left\{
 \begin{array}{ll}  \displaystyle
 \sqrt{x^{-1}-1}\arcsin\sqrt{x} & x\leq 1
 \\
 \displaystyle \frac{\sqrt{1-x^{-1}}}{2}\left[ \log\frac{1+\sqrt{1-x^{-1}}}
 {1-\sqrt{1-x^{-1}}}-i\pi \right]^2 \hspace{0.5cm} & x>1 \ .
\end{array} \right.
\end{eqnarray}

\vspace{1cm}

\def\theequation{B-\arabic{equation}}
\def\thesubsection{B}
\setcounter{equation}{0}
 \subsection{Mass corrections}
 \label{appendixB}

The MSSM higgses masses are, at one-loop for dominant top Yukawa (upper sign for $h$)
\medskip
\bea
\label{mh}
M_{h,H}^2&=&\frac{1}{2}\,\Big\{
m_A^2+m_Z^2+\delta\,m_Z^2\,\sin^2\beta
\mp \sqrt w\Big\}
\eea

\medskip\noindent
The mass of CP-odd Higgs is also modified by the effective operators
\medskip
\bea
m^2_{A}=\frac{2\,B\,\mu}{\sin 2\beta}\,
-\frac{2\,v^2}{\sin2\beta} \Big[\frac{c_0}{\Lambda 
}\,\mu\Big]
-\frac{v^4}{4}  \frac{\cos^2 2\beta}{\sin 2\beta}\,   
\Big[\frac{g_1^2c_1}{\Lambda^2}+ \frac{g_2^2c_2}{\Lambda^2}\Big]
+\cO\!\left(\!\frac{1}{\Lambda^3}\!\right)\label{mA}
\eea

\medskip\noindent
and $w=[(m_A^2-m_Z^2)\,\cos 2\beta+\delta\,m_Z^2\,\sin^2\beta]^2
+\sin^2 2\beta \,(m_A^2+m_Z^2)^2$.
Here $\delta$ is the  top/stop correction to the  Higgs potential, as in 
$\Delta V_h=(1/8)\,(g^2_1+g_2^2)\,\delta \,\vert h_u\vert^4$
where
\medskip
\bea
\delta  &\equiv&\frac{3\,h_{t}^{4}}{g^{2}\,\pi ^{2}\,}\bigg[\ln \frac{M_{\tilde{t}
}}{m_{t}}+\frac{X_{t}}{4}+\frac{1}{32\pi ^{2}}\,\Big(3\,h_{t}^{2}-16
\,g_{3}^{2}\Big)\Big(X_{t}+2\ln \frac{M_{\tilde{t}}}{m_{t}}\Big)\ln \frac{M_{
\tilde{t}}}{m_{t}}\bigg]
\nonumber\\[4pt]
X_{t} &\equiv & \frac{2\,(A_{t}-\mu \cot \beta )^{2}}{ M_{\tilde{t}}^{2}}
\,\,\Bigg(1- \frac{(A_{t}-\mu \cot \beta )^{2} }{ 12\,\,M_{\tilde{t}}^{2} }\,\Bigg). 
\end{eqnarray}

\medskip\noindent
with $M_{\tilde{t}}^{2}\equiv m_{\tilde{t}_{1}}m_{\tilde{t}_{2}}$, 
and $g_{3}$ is the QCD coupling.

The values of $\rho_{1,2}$ in eq.(\ref{deltam2}) are
\bea\label{md5}
 \rho_1& =& \,v^2\,\sin 2\beta\,\Big\{1\pm  \frac{(m_A^2+m_Z^2)}{\sqrt w}\Big\}
\\[6pt]
 \rho_2&=&\!\!
  \frac{v^4}{4\,\mu^2}\sin^2 2\beta
 \pm \frac{v^4}{\sqrt w} \Big\{-1+\frac{1}{2\mu^2} (m_A^2+m_Z^2) \sin^2 2\beta\Big\}
 \pm
 \frac{1}{w^{3/2}}\,(m_A^2+m_Z^2)^2\,v^4\,\sin^2  2\beta\nonumber
\eea
with the upper (lower) sign for $h$ ($H$).

\vspace{0.5cm}

\noindent
{\bf Acknowledgements:    }
The  work of D.M.G.  was supported by a grant of the Romanian National  Authority for
Scientific Research (CNCS-UEFISCDI) under  project number PN-II-ID-PCE-2011-3-0607.
The work of H.M.L. was supported in part by Basic Science Research Program through
the National Research Foundation of Korea (NRF) funded by the Ministry of Education,
Science and Technology (NRF-2016R1A2B4008759).

\end{document}